\documentclass[10pt,english,a4paper]{article}
\usepackage{a4wide}
\usepackage{color}
\usepackage{array}
\usepackage{float}
\usepackage{textcomp}
\usepackage{multirow}
\usepackage{amsmath}
\usepackage{amssymb}
\usepackage{graphicx}

\newcommand{\lyxmathsym}[1]{\ifmmode\begingroup\def\b@ld{bold}
  \text{\ifx\math@version\b@ld\bfseries\fi#1}\endgroup\else#1\fi}

\usepackage{babel}
\begin{document}
%\include{\string"MSc Thesis FINAL - Abstract redacted\string"}

%%%%%%%%%%%%%%%%%%%%%%%%%%%%%% User specified LaTeX commands.
\title{Modelling of dependence in high-dimensional financial time series by cluster-derived canonical vines}
\author{David Walsh-Jones\footnote{david@walsh-jones.com}, Daniel Jones\footnote{daniel.jones@maths.ox.ac.uk, Mathematical Institute, Oxford University, Andrew Wiles Building, Woodstock Road, Oxford, OX2 6GG, United Kingdom},
Christoph Reisinger\footnote{christoph.reisinger@maths.ox.ac.uk, Mathematical Institute, Oxford University, Andrew Wiles Building, Woodstock Road, Oxford, OX2 6GG, United Kingdom}}
\maketitle

%\tableofcontents{}

%\include{\string"Paper - Introduction\string"}\include{\string"Paper - StatVine\string"}\include{\string"Paper - Results\string"}

\begin{abstract}
\noindent We extend existing models in the financial literature by
introducing a cluster-derived canonical vine (CDCV) copula model for
capturing high dimensional dependence between financial time series.
This model utilises a simplified market-sector vine copula framework
similar to those introduced by Heinen and Valdesogo (2008) and 
Brechmann and Czado (2013),
which can be applied by conditioning asset time series on a market-sector
hierarchy of indexes. While this has been shown by the aforementioned
authors to control the excessive parameterisation of vine copulas
in high dimensions, their models have relied on the provision of externally
sourced market and sector indexes, limiting their wider applicability
due to the imposition of restrictions on the number and composition
of such sectors. By implementing the CDCV model, we demonstrate that
such reliance on external indexes is redundant as we can achieve equivalent
or improved performance by deriving a hierarchy of indexes directly
from a clustering of the asset time series, thus abstracting the modelling
process from the underlying data.\end{abstract}
\maketitle

%\nopagebreak

\section{Introduction\label{sec:Introduction}}

This paper introduces a new model for capturing high dimensional dependence
which we term the cluster-derived canonical vine (CDCV) copula model,
with a direct application to the practical modelling of large portfolios
of financial assets. Whilst the implementation of such advanced dependence
models is infrequent in the financial industry, more basic dependence
models are none-the-less heavily utilised. The ability to describe
the behaviour of a given financial variable in terms of other financial
variables enables us to both make use of the proliferation of data that
is available in the market and to derive proxies for financial variables
when data is not available. Moreover, when we consider the behaviour
of financial variables such as basket options, equity portfolios or
complex credit products that are directly dependent on their constituent
variables, we clearly require a method of capturing not just the marginal
behaviour of the constituents, but also the evolution of the dependence
structure between the constituents.\\

One of the more basic approaches to capturing such multivariate dependence
is the multivariate copula. First introduced by \cite{Sklar1959}
as a statistical tool, the copula decomposes a given multivariate
distribution into a dependence structure and a set of marginal distributions.
Multivariate copulas have arguably become an industry standard for
capturing dependence despite the negative press that the Gaussian
copula based Default Correlation model of \cite{Li1999} garnered
in the wake of the 2008 financial downturn (see \cite{Salmon2009}).
Due to its ability to capture stylised features of financial variables
such as fat tails, the Student's-t copula in particular is commonly
utilised. However, despite their widespread use, such parametric multivariate
copulas still present a level of inflexibility in that they are essentially
``one-size-fits-all'' and may not fully capture the nuances of
a given multivariate dependence structure.\\

This inflexibility has been addressed in the academic sphere by the
introduction of highly parametrised vine copulas (see \cite{Joe1996a,Bedford2002}),
which decompose the copula dependence structure into a collection
of trees containing bivariate copulas. Vine copulas enable the modeller
to select different bivariate copulas to represent the dependence
between different pairs of variables. This has the obvious advantage
of more accurately capturing complex and heterogeneous dependence
structures, and provides access to the much broader range of bivariate
copulas that exist for capturing features such as tail dependence.
Recent growth in the vine copula literature can be traced to the paper
``Pair-Copula Constructions of Multiple Dependence'' \cite{Aas2006}
which built upon the work of \cite{Joe1996a} and \cite{Bedford2002}
by firstly bringing their introduction of the vine copula to the forefront
of the literature and secondly, by illustrating how a vine copula
model could be constructed and fitted to data, given a set of marginal
distributions and a cascade of conditional pair copulas\textcolor{black}{.
}However, the practical application of vine copulas has been limited
to relatively low dimensional problems due to the need to fit the
parameters of as many as $m(m-1)/2$ bivariate copulas in an $m$-dimensional
model. Even with the latest computational technology the model fitting
process quickly becomes infeasible for these models in higher dimensions.
\\

To overcome this curse of dimensionality, a number of techniques have
been proposed. The most straightforward approach is that of vine simplification
or truncation, as described by \cite{Heinen2008}, \cite{Kurowicka2011}
and \cite{Brechmann2013a}, among others. This approach essentially
approximates the vine copula, by taking advantage of the \emph{de
minimus} contribution of later vine trees to the modelled dependence
structure. Secondly, the class of Market Sector Vine Copula models
(such as the CAVA model of \cite{Heinen2008} and the RVMS model
of \cite{Brechmann2013a}) aims to reduce the implementation cost
of vine copulas significantly via the introduction of a pre-existing
market-sector index hierarchy (such as the S\&P500) upon which elements
may be conditioned given simplifying assumptions regarding inter-sector
dependence. By conditioning asset time series on these index time
series, such models have enabled the flexibility of vine copulas to
be applied to portfolios of much higher dimensions by limiting the
number of trees that need to be fitted to achieve a fixed level of
model accuracy. Finally, recent research by \cite{Brechmann2013b,Brechmann2014233,Krupskii:2013:FCM:2501262.2501499,year,Joe2014}
has sought to develop hierarchical vine models that are not reliant
on externally sourced hierarchies, in a similar spirit to our own
research. For example, \cite{Brechmann2014233} use factor analysis
to develop latent factors upon which all elements are then conditioned
before utilising a truncated R-Vine copula to capture the remaining
idiosyncratic dependence between elements. The approach of \cite{year}
similarly uses factor analysis to derive the root nodes of C-Vine
copula trees. The authors do not seek to segment or cluster the population
of elements in the style of market sector vine copulas, but rather
to utilse underlying factors common to all elements. More recently,
\cite{Krupskii:2013:FCM:2501262.2501499} and \cite{Joe2014} have
proposed and then defined the Bi-factor copula model which can be
used when we have many variables which are divided into groups, making
the natural step of combining the market-sector hierarchy of \cite{Heinen2008,Brechmann2013a}
with the derivation of latent factors for both the market and the
sector groups.\\

Our research and development of the CDCV model is also motivated by
the Market Sector Vine Copula models of \cite{Heinen2008} and \cite{Brechmann2013a},
which focus on illustrative examples utilising specific externally-introduced
market-sector hierarchies. It is not immediately clear to what extent
these models can be applied to other data sets; whether the model
performance varies based on the external indexes used; whether the
size, number or composition of the sectors impacts model performance;
whether dependence structures and model performance vary through time,
or even whether it is always appropriate or possible to use such external
indexes. It is this class of models that we extend via the introduction
of the CDCV model, which mirrors the recently proposed Bi-factor copula
model of \cite{Joe2014} by replacing the externally sourced S\&P500
and Euro Stoxx 50 indexes of the CAVA and RVMS models respectively
with derived variables. An additional feature of the CDCV model is
that we can apply this market-sector hierarchical structure to any
data set, irrespective of whether the data is already grouped into
obvious segments or clusters. We do so by applying clustering and
index construction methodologies to the data, which allows the resulting
market-cluster hierarchical structure to vary in time and allows variables
to move between clusters. As such, the derived cluster indexes of
the CDCV model represent discrete dynamic clusters of elements that
may be considered analogous to sub-portfolios or trading books in
a financial context. The CDCV approach is thus additive in principle,
in that as larger pools of underlying elements are considered, the
derived indexes can be combined and re-used as necessary providing
consistent index construction methodologies are used. This leads us
to question the practical limitations of such factor-copula models,
which we begin to address in Section \ref{sec:Analysis,-Results-and}.
\\

In the following section, we formally introduce the CDCV model, outlining
the fundamental clustering and index creation steps while referring
the reader to the Appendix for details of the model fitting process,
performed using the Inference Functions for Margins (IFM) method of
\cite{Joe1996}. In Section \ref{sec:Analysis,-Results-and} we provide
an empirical analysis of the CDCV model's performance against an equivalent
(fixed hierarchy) market sector model of the CAVA-type proposed by
\cite{Heinen2008}. In this section we demonstrate that such models
need not rely on an external hierarchy and that equivalent or better
performance can be obtained by conditioning assets on indexes derived
directly from the underlying data. We also extend the analysis of
\cite{Heinen2008,Brechmann2013a,Joe2014,Krupskii:2013:FCM:2501262.2501499}
by demonstrating that the composition of Market Sector Vine Copula
models has a material impact on model performance and that model performance
is time-dependent. Finally, we conclude and discuss areas for further
research in Section \ref{sec:Discussion-of-Findings}.

\section{The CDCV Model\label{sec:The-StatVine-Model}}

\begin{figure}[H]
\centering
%\begin{center}
%\textbf
{\includegraphics[width=0.8\columnwidth]{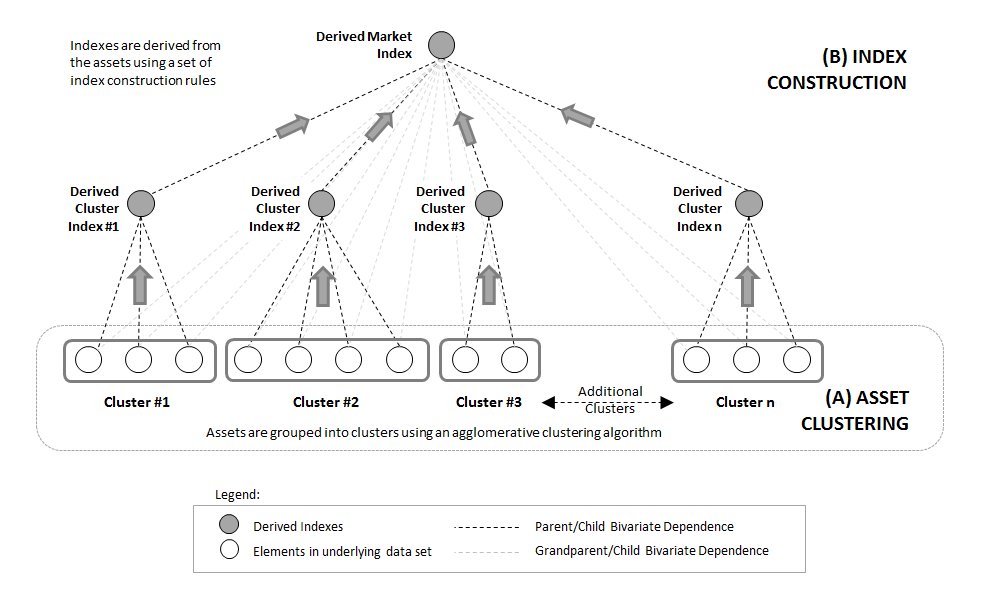}}
%\vspace{-1 cm}
%\end{center}
\caption{Diagrammatic representation of the
CDCV model's derived hierarchical structure, obtained by grouping
assets into $n$ clusters and then constructing indexes for each.
The market index can then be constructed from the cluster indexes
or directly from all assets.}
\label{fig:1}
\end{figure}

We now formally introduce in detail the proposed cluster-derived canonical
vine (CDCV) copula model, depicted in Figure~\ref{fig:1}. By
deriving indexes rather than utilising externally sourced indexes,
we mirror the Bi-factor copula model approach of \cite{Joe2014}
but extend the model's applicability to arbitrary data sets via additional
clustering and index derivation steps. This extension is important
in practice as we may find that there are many natural groupings of
the variables and it may not be practical to fit all of them. We perform
conditioning as part of a C-vine fitting process upon indexes derived
from these asset clusters. We allow these clusters to evolve through
time subject to a set of configurable clustering rules (see Appendices
\ref{sub:Clustering-Rules--} and \ref{sub:Clustering-Rules---1})
and using a general clustering algorithm (see Appendix \ref{sub:Clustering-Algorithm--}).
As such, we may draw a parallel to the management of portfolios in
a financial setting.

%\begin{center}
%\end{center}

In order to implement this model we face two primary challenges; firstly
how to group or cluster the assets to maximise the dependence captured
by the model, and secondly how to use these groupings to derive optimal
sector and market indexes. Figure~\ref{fig:1} illustrates these
challenges in the context of our proposed CDCV model, for which we
will utilise the same hierarchical C-Vine decomposition as the CAVA
model of \cite{Heinen2008} to enable us to compare performance against
that model in Section \ref{sec:Analysis,-Results-and}. This decomposition
can be defined in this more general setting as\\
\begin{equation}
f_{CDCV}\left(r_{M},\, r_{S_{1}},...,\, r_{S_{E}},\, r_{1}^{S_{1}},...,\, r_{Z^{1}}^{S_{1}},......,\, r_{1}^{S_{E}},...,\, r_{Z^{E}}^{S_{E}}\right)=\bar{f}\cdot\bar{c}_{M,S}\cdot\bar{c}_{M,A}\cdot\bar{c}_{S,A\mid M}\cdot\bar{c}_{A}\,,\label{eq:StatVine decomp}
\end{equation}
where $r_{M}$ is the market index return, $r_{S_{i}}$ are the $1\leq i\leq E$
cluster index returns, and $r_{j}^{S_{i}}$ are the $1\leq j\leq Z_{i}$
asset returns associated with cluster $i$. The marginals appear in
\begin{equation}
\begin{array}{ccl}
\bar{f} & = & f\left(r_{M}\right)\cdot f\left(r_{S_{1}}\right)\cdot...\cdot f\left(r_{S_{E}}\right)\cdot\left[f\left(r_{1}^{S_{1}}\right)\cdot...\cdot f\left(r_{Z_{1}}^{S_{1}}\right)\right]\cdot...\cdot\left[f\left(r_{1}^{S_{E}}\right)\cdot...\cdot f\left(r_{Z_{E}}^{S_{E}}\right)\right]\end{array}\label{eq:HV Marg-1-1}.
\end{equation}
The unconditional copulas between the market index and the sector indexes are given by
\begin{equation}
\begin{array}{ccl}
\bar{c}_{M,S} & = & c_{r_{M},r_{S_{1}}}\left[{\scriptstyle F\left(r_{M}\right),F\left(r_{S_{1}}\right)}\right]\cdot...\cdot c_{r_{M},r_{S_{E}}}\left[{\scriptstyle F\left(r_{M}\right),F\left(r_{S_{E}}\right)}\right]\end{array}\label{eq:HV Uncond-1-1}
\end{equation}
and the remaining unconditional copulas between the market index and the assets are
\begin{equation}
\begin{array}{cccl}
\bar{c}_{M,A} & = & \left[c_{r_{M},r_{1}^{S_{1}}}\left[{\scriptstyle F\left(r_{M}\right),F\left(r_{1}^{S_{1}}\right)}\right]\cdot...\cdot c_{r_{M},r_{Z_{1}}^{S_{1}}}\left[{\scriptstyle F\left(r_{M}\right),F\left(r_{Z_{1}}^{S_{1}}\right)}\right]\right] & \cdot...\cdot\\
 &  & \left[c_{r_{M},r_{1}^{S_{E}}}\left[{\scriptstyle F\left(r_{M}\right),F\left(r_{1}^{S_{E}}\right)}\right]\cdot...\cdot c_{r_{M},r_{Z_{E}}^{S_{E}}}\left[{\scriptstyle F\left(r_{M}\right),F\left(r_{Z_{E}}^{S_{E}}\right)}\right]\right] & \,,
\end{array}\label{eq:HV Uncond 2-1-1}
\end{equation}
where $c_{r_{M},r_{S_{j}}}$ denotes the bivariate copula between
the market index and the sector $j$ index. The CDCV model then captures
the dependence between each asset and its respective sector index
(conditioned upon the market index) via conditional copulas, as represented
by the $\bar{c}_{S,A\mid M}$ term in (\ref{eq:StatVine decomp}).
In the context of a C-Vine copula, the market index can thus be considered
to be the root node of the first tree, while the subsequent trees
select the sector indexes as their nodes. The ordering of these subsequent
trees is arbitrary due to the assumption of conditional independence
between cluster indexes and between cluster indexes and assets from other clusters. 
The $\bar{c}_{S,A\mid M}$ term is given as
\begin{equation}
\begin{array}{ccl}
\bar{c}_{S,A\mid M} & = & \left[c_{r_{S_{1}},r_{1}^{S_{1}}\mid r_{M}}\left[{\scriptstyle F\left(r_{S_{1}}\mid r_{M}\right),F\left(r_{1}^{S_{1}}\mid r_{M}\right)}\right]\cdot...\cdot c_{r_{S_{1}},r_{Z_{1}}^{S_{1}}\mid r_{M}}\left[{\scriptstyle F\left(r_{S_{1}}\mid r_{M}\right),F\left(r_{Z_{1}}^{S_{1}}\mid r_{M}\right)}\right]\right]\cdot...\cdot\\
 &  & \left[c_{r_{S_{E}},r_{1}^{S_{E}}\mid r_{M}}\left[{\scriptstyle F\left(r_{S_{E}}\mid r_{M}\right),F\left(r_{1}^{S_{E}}\mid r_{M}\right)}\right]\cdot...\cdot c_{r_{S_{E}},r_{Z_{E}}^{S_{E}}\mid r_{M}}\left[{\scriptstyle F\left(r_{S_{E}}\mid r_{M}\right),F\left(r_{Z_{E}}^{S_{E}}\mid r_{M}\right)}\right]\right]\,,
\end{array}\label{eq:HV cond-1-1}
\end{equation}
where $c_{r_{S_{j}},r_{i}^{S_{j}}\mid r_{M}}$ is the bivariate copula
between sector $j$ and asset $i$ within that sector, conditioned
on the market index. Finally, the CDCV model captures any remaining
idiosyncratic dependence with a multivariate copula, utilising the
technique of Joint Simplification (see \cite{Heinen2008}), where
a multivariate copula is applied between all assets, each conditioned
on the market index and on their associated sector index. This is
represented in the decomposition by the $\bar{c}_{A}$ term, given
as
\begin{equation}
\begin{array}{ccl}
\bar{c}_{A} & = & c_{r_{1}^{S_{1}}...r_{Z^{1}}^{S_{1}}......r_{1}^{S_{E}}...r_{Z^{E}}^{S_{E}}\mid r_{M},r_{S_{1}},...,r_{S_{E}}}\\
 &  & \left[{\scriptstyle F\left(r_{1}^{S_{1}}\mid r_{S_{1}},\, r_{M}\right),...,F\left(r_{Z_{1}}^{S_{1}}\mid r_{S_{1}},\, r_{M}\right),...,...,F\left(r_{1}^{S_{E}}\mid r_{S_{E}},\, r_{M}\right),...,F\left(r_{Z_{E}}^{S_{E}}\mid r_{S_{E}},\, r_{M}\right)}\right]\,.
\end{array}\label{eq:HV MVT-1}
\end{equation}
While we have chosen to develop the CDCV model using the more standardised
C-Vine specification used by the CAVA model of \cite{Heinen2008},
a secondary step (not taken here) would be to assess the relative
impact of our findings when applied to the more generalised R-Vine
modelling structure, as utilised by \cite{Brechmann2014233}.

\subsection{Dynamically Grouping Assets into Clusters\label{sub:Dynamically-Grouping-Assets}}

\begin{figure}[H]
\centering
\textbf{\includegraphics[scale=0.45]{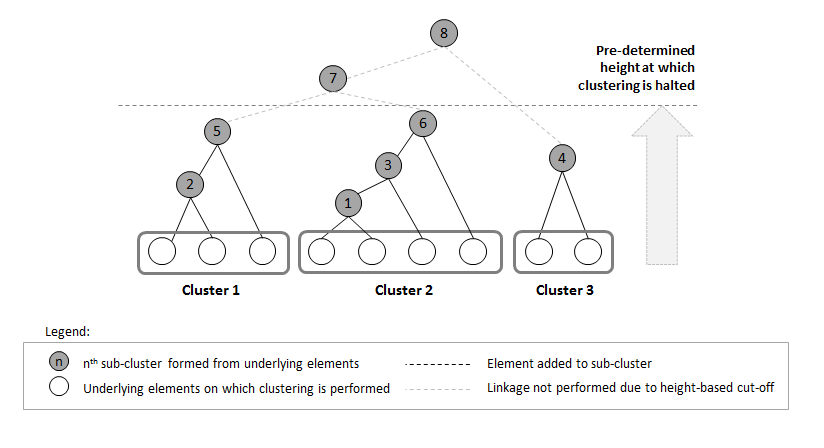}}
\caption{Diagrammatic representation of an
agglomerative clustering approach for 9 assets subject to a height-based
cut-off rule.}
\label{fig:2}
\end{figure}

As we choose to implement the same hierarchical C-Vine structure as
the CAVA model of \cite{Heinen2008}, we are interested in constructing
clusters that minimise the dependence between assets in different
clusters. While further work can be performed in this area to develop
algorithms that achieve such optimal clusterings and thus capture
the maximum possible dependence between assets, we will demonstrate
that even a heuristic approach to selecting clusters can result in
an improvement upon the existing sector-based approach of \cite{Heinen2008}.
For the purposes of our analysis we will consider only agglomerative
clustering methods, as visualised in Figure~\ref{fig:2}, which
seek to iteratively group assets until some predetermined condition
is met. These methods are less computationally intensive than divisive
clustering methods which start from one super-set cluster and iteratively
bifurcate the population(s) in each cluster. To develop clusters,
we calculate dissimilarity metrics for each pair of elements at each
iterative step in the process, as defined in Appendix \ref{sub:Clustering-Rules--}.
We then apply a clustering rule, known as a linkage criterion, at
each step to select which elements to join together into a cluster.
Examples of common linkage criteria are given in Appendix 
\ref{sub:Clustering-Rules---1}. Newly formed clusters then become
elements in the next step and may be selected for joining. To perform
this repeated joining of assets and clusters we may use a clustering
algorithm as provided in Appendix \ref{sub:Clustering-Algorithm--}.
Such an algorithm can then be controlled by the introduction of configurable
parameters into the algorithm or rule itself; for example, to ensure
a minimum cluster size, a fixed or varying number of clusters, and so on.

While the Euclidean distance is probably the most commonly used distance
metric in the clustering literature (see \cite{Shahid2009,everitt2001cluster,Brechmann2013b} for an overview),
we are more inclined to use rank correlation measures for this task
as we are looking to group time series data that demonstrate the most
dependence. In terms of the linkage criterion that we employ, the
choice is largely driven by the type of clusters we are looking to
produce. For example, the Average Linkage Criterion tends to join
clusters with small within-cluster variances and also tends to be
less affected by extreme values than many other methods. Alternatively,
the Complete Linkage Criterion can be significantly impacted by moderately
outlying values and is biased toward producing compact clusters of
approximately equal radius. In our analysis of the CDCV model we will
primarily choose to use an Adapted Single Linkage Criterion that we
have introduced, incorporating some additional rules not included
in the generic agglomerative clustering algorithm given in Appendix
\ref{sub:Clustering-Algorithm--}. This criterion is similar
to the standard Single Linkage Criterion, but it additionally limits
the size of any given cluster to a parametrised maximum number of
elements, limits the total number of clusters to a parametrised maximum
value and ignores potential joins where both elements are already
non-singleton clusters. This final restriction is implemented to avoid
chaining, which can be an issue with Single Linkage algorithms, where
each link covers a short distance but the most dissimilar elements
in a cluster may end up quite distant from each other. To this extent,
we can think of linkage criteria as not only a rule for deciding which
clusters to merge, but also as a means for introducing additional
conditions that provide greater control over the size, shape and composition
of the resulting clusters. While the dynamic clustering approach of
the CDCV model is clearly very intuitive, time-varying and a conceptual
improvement over the fixed sector clustering methods, it should be
noted that a further area for research remains to develop optimised
clustering methods.

\subsection{Deriving Hierarchical Indexes from the Assets\label{sub:Deriving-Hierarchical-Indexes}}

\begin{figure}[H]
\centering
\textbf{\includegraphics[scale=0.4]{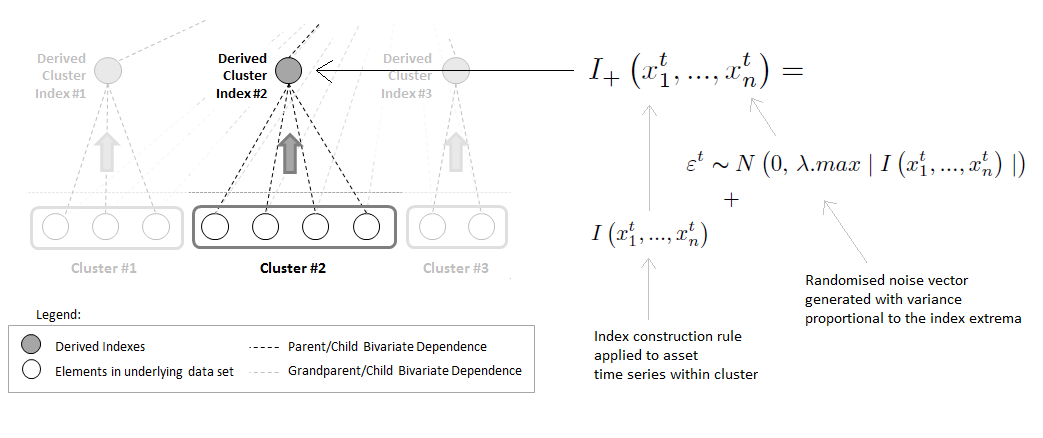}}
\caption{Diagrammatic representation of the
methodology by which cluster indexes are derived in the CDCV model.
A market index may then be derived either directly from the cluster
indexes or directly from the set of all assets.}
\label{fig:3}
\end{figure}

The assets that the CDCV model clusters into a particular grouping
in a given time step may not be immediately representable by an existing
index. We thus make use of a general index derivation methodology,
illustrated in Figure~\ref{fig:3}, from which we may construct
index(es) for each cluster to be used as latent variables in our model.
While there are many possible methods by which we may derive these
latent variables on which we will condition the assets, we will prefer
methods that provide relatively stable cluster indexes through time.

We outline in Appendix \ref{sub:Index-Construction--} a number
of basic but commonly used index construction rules (denoted $I\left(\cdot\right)$)
that are used in the financial industry, and combine these with
a normally distributed random variable ``noise'' vector given by

\begin{equation}
\varepsilon^{t}\thicksim N\left(0,\,\lambda \cdot \max\mid I\left(x_{1}^{t},...,x_{n}^{t}\right)\mid\right)\,,\label{eq:NOISE}
\end{equation}
where $\Upsilon=\frac{1}{\lambda}$ is a noise parameter that we use
to adjust the scale of the perturbations to be introduced. This enables
us to define our noise-adjusted index in each time step as

\begin{equation}
I_{+}\left(x_{1}^{t},...,x_{n}^{t}\right)=I\left(x_{1}^{t},...,x_{n}^{t}\right) + \varepsilon^{t}\,.\label{eq:noise}
\end{equation}
This noise term remediates an issue that arises from the process of
constructing indexes directly from a small number of asset time series
and then conditioning those time series on the resultant index. When
cluster sizes are small, we may end up introducing artificially high
levels of negative rank correlation into our model due to the index
representing too perfectly a median path between two asset time series.
As we will show in Section \ref{sec:Analysis,-Results-and}, this
noise term is sufficient to dampen the negative rank correlations
generated, while still capturing efficiently the positive dependence
in the underlying asset time series data. While the optimisation of
such index constructions is another area for further research, we
will demonstrate in Section \ref{sec:Analysis,-Results-and} that
with only minimal attention to this problem we are able to construct
sufficiently good indexes to obtain model fitting results that outperform
an equivalent model utilising the CAVA model's structure.

\subsection{Implementing the CDCV Model\label{sub:Implementing-the-StatVine}}

To implement the CDCV model as defined in this paper we have built
up a modelling structure and test framework using the statistical
programming language R. We have updated the algorithms described by
\cite{Heinen2008} to provide Inference Functions for Margins (see \cite{Joe1996})
model fitting and simulation algorithms for the CDCV model, which
we provide in Appendix \ref{sub:StatVine-Model-fitting-Algorithm}
and \ref{sub:StatVine-Simulation-Algorithm}. In these algorithms
we choose between Normal, Student's-t and Skew Student's-t marginal
distributions using the Akaike Information Criterion (AIC, as per \cite{Aike1974}),
to ensure that we can capture characteristics of financial asset return
time series such as excess skew and kurtosis. We also restrict ourselves
to homoscedastic marginal distributions, in line with \cite{Low2013}
who observe that the introduction of GARCH-type marginals had no noticeable
impact on the results of their vine-copula focused portfolio optimisation
analysis. Once we have fitted the marginal distributions, we transform
the marginal data to the unit hypercube. For each bivariate combination
of asset plus market index, we then maximise the bivariate log-likelihood
of selected bivariate copula families, given in generality by \cite{cherubini2004copula}
as
\begin{equation}
l(\Theta;\, x_{1},x_{2})=\overset{m}{\underset{i=1}{\sum}}\log\,\bar{C}_{\Theta}\left(F_{1}\left(x_{i,1}\right),F_{2}\left(x_{i,2}\right)\right)-\overset{m}{\underset{i=1}{\sum}}\overset{2}{\underset{j=1}{\sum}}\log\, f_{j}\left(x_{i,j}\right)\,,\label{eq: log lik for copula fitting}
\end{equation}
where $\bar{C}_{\Theta}$ is the copula density defined for each trialled
copula type. The set of marginal distributions is
\begin{equation}
\hat{\Omega}=\left\{ F_{1}\left(x_{1};\,\hat{z}_{1}\right),F_{2}\left(x_{2};\,\hat{z}_{2}\right),...,F_{m}\left(x_{m};\,\hat{z}_{m}\right)\right\} \,,\label{eq:Marg Distrib Pop}
\end{equation}
and the resultant set of estimated marginal densities is
\begin{equation}
\hat{\Psi}=\left\{ f_{1}\left(x_{1};\,\hat{z}_{1}\right),f_{2}\left(x_{2};\,\hat{z}_{2}\right),...,f_{n}\left(x_{m};\,\hat{z}_{m}\right)\right\} \,,\label{eq:Marginal Density Pop}
\end{equation}
where $\hat{z}=\left\{ \hat{z}_{1},...,\hat{z}_{m}\right\} $ is the
set of estimated marginal parameters. Note that the second term of
(\ref{eq: log lik for copula fitting}) does not depend on the copula
parameter(s), and thus for the IFM approach we need only maximise
the first term. The resulting log-likelihoods for the Gaussian, Student's-t,
Clayton and Frank copula families enable selection of the best fitting
bivariate copula, again by AIC. However, model-fitting a C-Vine copula
also requires us to apply an $h$-function (\ref{eq: h function-1})
after each bivariate copula in the vine is fitted, in order to transform
the sample data used to fit the copula into sample data which is additionally
conditioned on the current root node, to be used in fitting the conditional
bivariate copulas in the next tree. These $h$-functions are a simplified
form of the vine copula conditional distribution function, 
given by \cite{Joe1996a} and \cite{Heinen2008} as
\begin{equation}
F_{n\mid n-1}\left(x_{n}\mid x_{n-1}\right)=\frac{\partial C_{n,(n-1)_{j}\mid(n-1)_{-j}}\left[F\left(x_{n}\mid x_{(n-1)_{-j}}\right),\, F\left(x_{(n-1)_{j}}\mid x_{(n-1)_{-j}}\right)\right]}{\partial F\left(x_{(n-1)_{j}}\mid x_{(n-1)_{-j}}\right)}\,,\label{eq: Cond Distrib Function}
\end{equation}
where for notational convenience $c_{-j}$ is defined as the vector
$c$ but without component $j$, and where $n-1$ can be taken to
represent a string of previously conditioned variable indexes up to
that value. Following \cite{Joe1996a}, the $h$-function may be written as
\begin{eqnarray}
h\left(x_{n},x_{n-1},\theta\right) & = & F_{n\mid n-1}\left(x_{n}\mid x_{n-1}\right)\nonumber \\
 & = & \frac{\partial C_{\theta_{n,(n-1)}}\left[F\left(x_{n}\right),\, F\left(x_{n-1}\right)\right]}{\partial F\left(x_{n-1}\right)}, \label{eq: h function-1}
\end{eqnarray}
where $F(\cdot)$ represent marginal distributions that have already been conditioned successively on root 
nodes from earlier trees. In (\ref{eq: h function-1}), $x_{n}$ and $x_{n-1}$ are univariate 
(and in practice, uniform) and are defined for each copula family (see \cite{Heinen2008} for
a table). Furthermore, $\theta$ represents the copula
parameter(s) for the copula family fitted between the $n^{th}$ and
$(n-1)^{th}$ nodes (after conditioning on nodes $1$ to $n-2$).
We can generalise this iterative conditioning and express the $n$-dimensional
C-Vine copula density per \cite{Aas2006,Heinen2008} as
\begin{equation}
c_{12...n}\left[F_{1}\left(x_{1}\right),F_{2}\left(x_{2}\right),...,F_{n}\left(x_{n}\right)\right]=\prod_{j=1}^{n-1}\prod_{k=1}^{n-j}c_{j,j+k\mid1,...,j-1}\left[{\scriptstyle F\left(x_{j}\mid x_{1},...x_{j-1}\right),\, F\left(x_{j+k}\mid x_{1},...x_{j-1}\right)}\right]\,,\label{eq:C-Vine Copula Density-1}
\end{equation}
where $j=1$ implies an absence of conditioning. Equivalently, we
can express the C-Vine copula's log-likelihood function as
\begin{equation}
L\left(x_{1},...,x_{n};\theta\right)=\sum_{j=1}^{n-1}\sum_{k=1}^{n-j}\sum_{t=1}^{\tau}\log\left(c_{j,j+k\mid1,...,j-1}\left[{\scriptstyle F\left(x_{j,t}\mid x_{1,t},...x_{j-1,t}\right),\, F\left(x_{j+k,t}\mid x_{1,t},...x_{j-1,t}\right)}\right]\right)\,,\label{eq:Log Lik vine decomp-1}
\end{equation}
where $\theta$ is the set of the C-Vine's parameters and we assume
for simplicity that we are fitting time series containing $\tau$
independent observations. Equation (\ref{eq:Log Lik vine decomp-1})
illustrates that the log-likelihood of a C-Vine can be decomposed
into a sum of bivariate log-likelihoods. Given this, we may implement
an algorithm that initially fits unconditional bivariate copulas in
each tree of the vine by maximising their respective log-likelihoods,
and then accounts for the necessary conditioning in subsequent trees
by iteratively transforming the observed data using $h$-functions
per (\ref{eq: h function-1}). We provide in Appendix \ref{sub:C-Vine-Model-fitting-Algorithm}
pseudo-code for a general C-Vine copula fitting algorithm that utilises
these $h$-functions and selects copulas according to their AIC statistic,
based on the algorithms provided by \cite{Aas2006}. 

\begin{figure}[H]
\centering
\textbf{\includegraphics[scale=0.4]{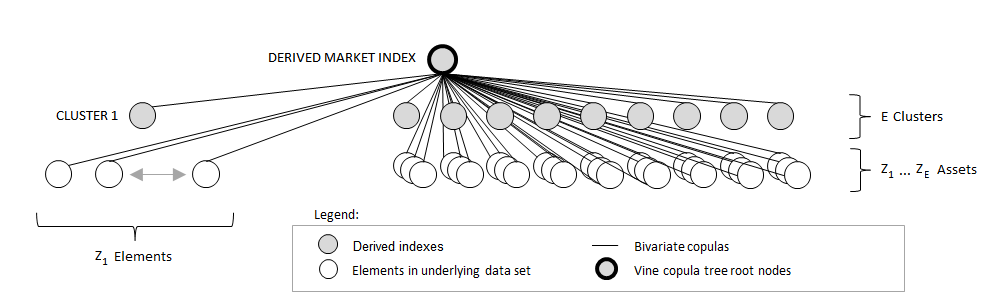}}
\caption{Diagrammatic representation of the
first tree in the CDCV model.}
\label{fig:4}
\end{figure}
A fitting algorithm for the CDCV model is also provided in Appendix
\ref{sub:StatVine-Model-fitting-Algorithm}, which loops through
each cluster, fitting firstly the cluster index to market index unconditional
copula and secondly the asset to market index unconditional copulas.
This process fits the first C-Vine tree of the CDCV model, as illustrated
in Figure~\ref{fig:4}. In doing so, we transform the cluster
index and asset time series using the fitted parameters and appropriate
$h$-function for the AIC-selected copula family.

\begin{figure}[H]
\centering
\textbf{\includegraphics[scale=0.4]{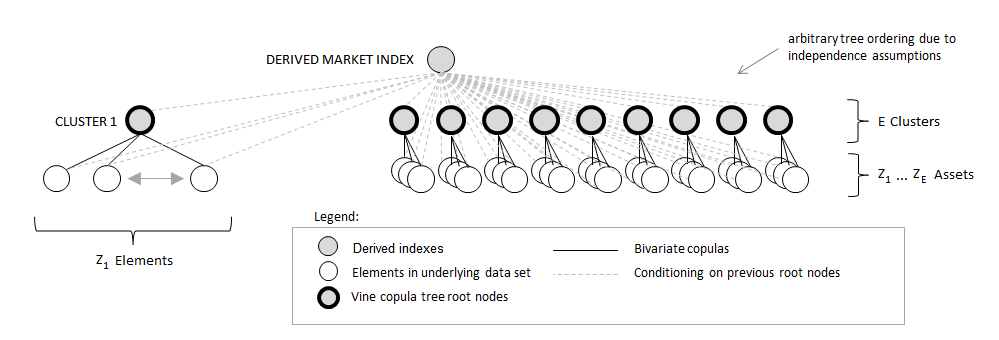}}
\caption{Diagrammatic representation of the
subsequent trees in the CDCV model, with cluster indexes as root nodes.}
\label{fig:5}
\end{figure}
The CDCV model's fitting algorithm is a simple extension of the C-Vine
algorithm, based on the method of \cite{Heinen2008}. The primary
differences between the CDCV and C-Vine fitting algorithms are that
the CDCV algorithm fits a multivariate copula after fitting a specified
number of trees (i.e., it is a simplified C-Vine), it incorporates
the concept of clustering and it incorporates independence assumptions
between elements and indexes from other clusters. After fitting the
first tree of the CDCV model, we then fit a conditional copula between
each asset and its associated cluster index (i.e., conditional upon
the market index), as illustrated in Figure~\ref{fig:5}.

\begin{figure}[H]
\centering
\textbf{\includegraphics[scale=0.4]{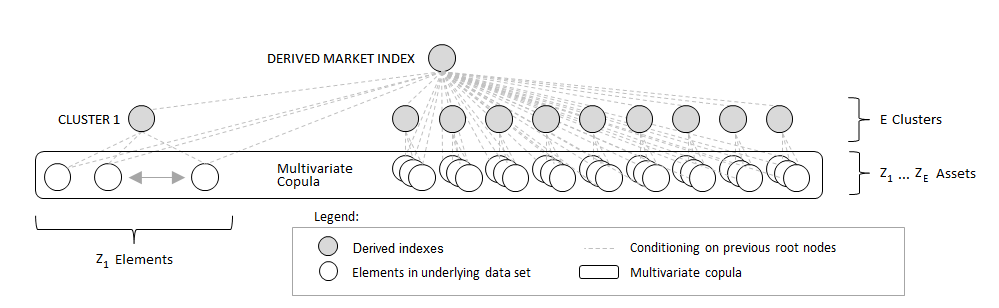}}
\caption{Diagrammatic representation of the
CDCV model's jointly-simplifying multivariate copula.}
\label{fig:6}
\end{figure}
Finally, we fit a Student's-t or Gaussian multivariate copula to the
conditioned assets, as illustrated in Figure~\ref{fig:6}. A
similar adaption of the C-Vine simulation algorithm (given in Appendix
\ref{sub:C-Vine-Simulation-Algorithm}) enables us to easily
simulate from the CDCV model, as detailed in Appendix \ref{sub:StatVine-Simulation-Algorithm}.

\section{Analysis, Results and Conclusions\label{sec:Analysis,-Results-and}}

In order to demonstrate that our CDCV model is capable of providing
improved results over equivalent fixed-hierarchy models in the literature,
we implement first a version of the Heinen \& Valdesogo CAVA model
selecting between the marginal distributions, bivariate copulas and
multivariate copulas described in Section \ref{sub:Implementing-the-StatVine}.
We then implement the CDCV model by replacing the externally sourced
S\&P 500 indexes with the CDCV model's derived indexes as outlined
in Section \ref{sub:Deriving-Hierarchical-Indexes}, before then also
relaxing the fixed clustering structure and allowing it to vary through
time based on the clustering methodology detailed in Section \ref{sub:Dynamically-Grouping-Assets}.

\subsection{Data\label{sub:Data}}

To test both the CDCV and CAVA implementations with clusters of varying
size, we select the S\&P 500 market and 10 industry sector indexes,
plus 62 of the 95 assets that \cite{Heinen2008} analysed.

\begin{table}[H]
\centering
\begin{tabular*}{35pc}{@{\extracolsep{\fill}}c|ccccc|ccccc}
\hline 
\textcolor{black}{\tiny{H\&V Sector}} & \multicolumn{5}{c|}{\textcolor{black}{\tiny{Largest 5 Stocks by Market Cap June 2008}}} & \multicolumn{5}{c}{\textcolor{black}{\tiny{Smallest 5 Stocks by Market Cap June 2008}}} \\
\hline 
\textcolor{black}{\tiny{ENERGY}} & \textcolor{black}{\tiny{XOM}} & \textcolor{black}{\tiny{CVX }} & \textcolor{black}{\tiny{COP}} & \textcolor{black}{\tiny{SLB}} & \textcolor{black}{\tiny{OXY}} & \textcolor{black}{\tiny{RDC}} & \textcolor{black}{\tiny{TSO}} &  &  &  \\
\hline 
\textcolor{black}{\tiny{INDUSTRIAL}} & \textcolor{black}{\tiny{GE}} & \textcolor{black}{\tiny{UTX}} & \textcolor{black}{\tiny{BA}} & \textcolor{black}{\tiny{MMM}} & \textcolor{black}{\tiny{CAT}} & \textcolor{black}{\tiny{PLL}} & \textcolor{black}{\tiny{R}} & \textcolor{black}{\tiny{CTAS}} & \textcolor{black}{\tiny{RHI}} &  \\
\hline 
\textcolor{black}{\tiny{HEALTH}} & \textcolor{black}{\tiny{JNJ}} & \textcolor{black}{\tiny{PFE}} & \textcolor{black}{\tiny{MRK}} & \textcolor{black}{\tiny{ABT}} &  & \textcolor{black}{\tiny{PKI}} & \textcolor{black}{\tiny{THC}} &  &  &  \\
\hline 
\textcolor{black}{\tiny{FINANCIAL}} & \textcolor{black}{\tiny{BAC}} & \textcolor{black}{\tiny{JPM}} & \textcolor{black}{\tiny{C}} & \textcolor{black}{\tiny{AIG}} & \textcolor{black}{\tiny{WFC}} & \textcolor{black}{\tiny{HBAN}} &  &  &  &  \\
\hline 
\textcolor{black}{\tiny{UTILITIES}} & \textcolor{black}{\tiny{EXC}} & \textcolor{black}{\tiny{SO}} & \textcolor{black}{\tiny{D}} & \textcolor{black}{\tiny{DUK}} &  & \textcolor{black}{\tiny{TEG}} & \textcolor{black}{\tiny{TE}} & \textcolor{black}{\tiny{PNW}} & \textcolor{black}{\tiny{CMS}} & \textcolor{black}{\tiny{GAS}} \\
\hline 
\textcolor{black}{\tiny{MATERIALS}} & \textcolor{black}{\tiny{DD}} & \textcolor{black}{\tiny{DOW}} & \textcolor{black}{\tiny{AA}} & \textcolor{black}{\tiny{PX}} & \textcolor{black}{\tiny{NUE}} & \textcolor{black}{\tiny{IFF}} & \textcolor{black}{\tiny{BMS}} &  &  &  \\
\hline 
\textcolor{black}{\tiny{CONS DISCR}} & \textcolor{black}{\tiny{MCD}} & \textcolor{black}{\tiny{CMCSA}} & \textcolor{black}{\tiny{DIS}} & \textcolor{black}{\tiny{HD}} &  &  &  &  &  &  \\
\hline 
\textcolor{black}{\tiny{CONS STAP}} & \textcolor{black}{\tiny{PG}} & \textcolor{black}{\tiny{WMT}} & \textcolor{black}{\tiny{KO}} & \textcolor{black}{\tiny{PEP}} & \textcolor{black}{\tiny{CVS}} & \textcolor{black}{\tiny{BF.B}} &  &  &  &  \\
\hline 
\textcolor{black}{\tiny{IT}} & \textcolor{black}{\tiny{MSFT}} & \textcolor{black}{\tiny{IBM}} & \textcolor{black}{\tiny{AAPL}} & \textcolor{black}{\tiny{CSCO}} & \textcolor{black}{\tiny{INTC}} &  &  &  &  &  \\
\hline 
\textcolor{black}{\tiny{TELECOM}} & \textcolor{black}{\tiny{T}} & \textcolor{black}{\tiny{VZ}} & \textcolor{black}{\tiny{CTL}} &  &  & \multicolumn{5}{c}{} \\
\hline 
\end{tabular*}

%\bigskip{}
%\medskip{}

\caption{{Details of the assets we use for analysing
the CDCV model. These are 62 of the 95 assets used by Heinen \& Valdesogo
to test their CAVA model, and provide us with variation in the number
of stocks from each industry.}}
\label{tab:1}
\end{table}

For these stocks and indexes, we obtained from Bloomberg daily return
values between 1st January 2005 and 18th December 2008 to analyse
performance both prior to and during the recent financial crisis. 

\begin{figure}[H]
\centering
\textbf{\includegraphics[scale=0.33]{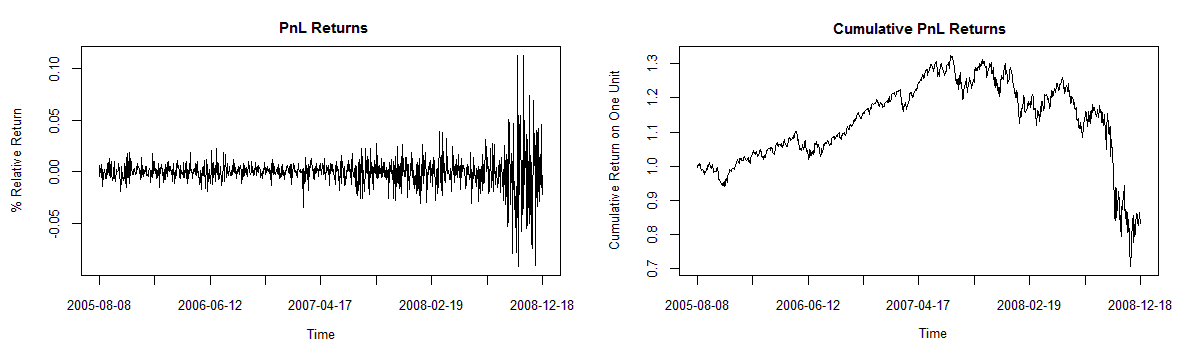}}
\caption{Daily and cumulative relative returns
PnL of an equally weighted portfolio of the 62 assets in Table 1,
between 8th August 2005 and 18th December 2008 (i.e., excluding the
initial 150 days learning period).}
\label{fig:7}
\end{figure}

\begin{figure}[H]
\centering
\includegraphics[scale=0.48]{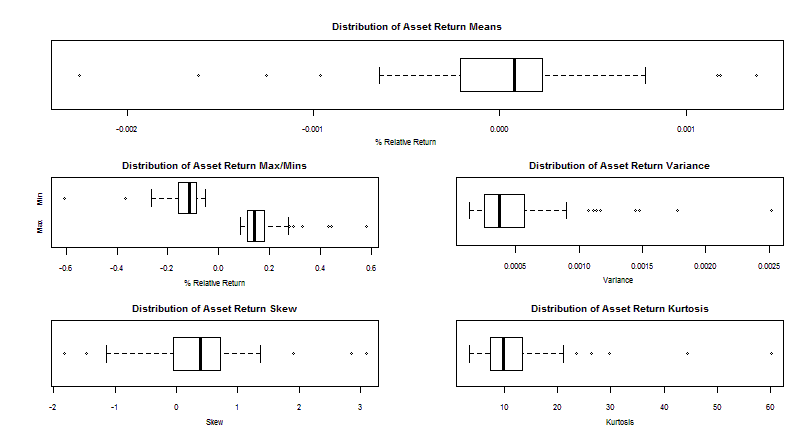}
\caption{Distribution of the 62 %marginal distributions'
relative asset return means, maxima, minima, variance, skew and
kurtosis, where the statistics of each marginal distribution are obtained
directly from the data between 8th August 2005 and the 18th December
2008 (i.e., excluding the initial 150 days learning period).}
\label{fig:8}
\end{figure}

This data is illustrated in Figure~\ref{fig:7}, showing the
daily and cumulative relative returns for an evenly weighted portfolio
of the 62 marginals, with an average daily return of $0.00002$\%
and a variance of $0.00052$\%. The distribution of these asset return
means is illustrated in Figure~\ref{fig:8}, and clearly shows
that the presence of negative skew in the asset returns. We also note
that these 62 marginal distributions have a mean kurtosis of $12.48$,
and a minimum kurtosis of $3.416$, which strongly indicates that
we have non-Gaussian marginals.

\begin{figure}[H]
\centering
\includegraphics[scale=0.6]{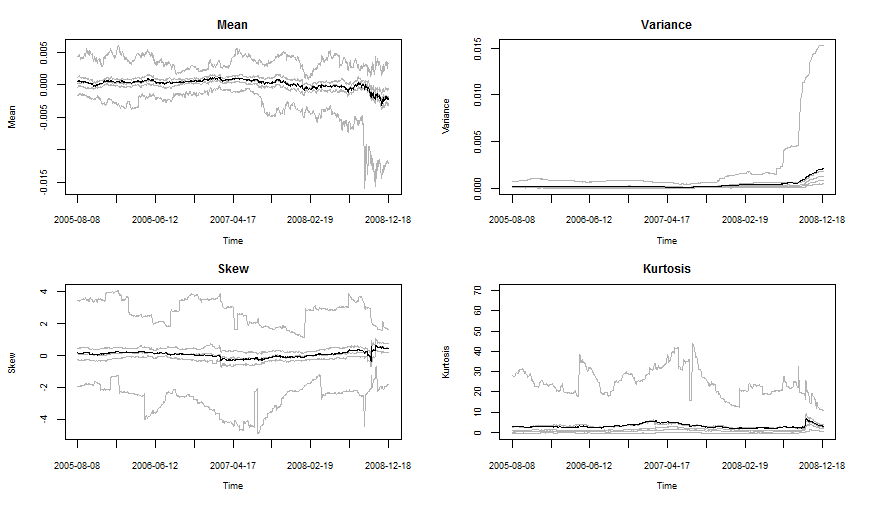}
\caption{Mean (black) and quantiles (grey)
q1, q25, q50, q75, q99 of the 62 marginals' relative asset return
mean, variance, skew and kurtosis statistics through time, based on
a 150 day rolling learning period.}
\label{fig:9}
\end{figure}

In the following analysis, we are also interested in the time-varying
performance of the CDCV and CAVA models; an aspect of market sector
model performance not directly addressed by either \cite{Heinen2008}
or \cite{Brechmann2013a}, or the related literature. To support
this analysis, we illustrate in Figure~\ref{fig:9} the time-dependent
variation of the marginal data statistics from Figure~\ref{fig:8}.
Of particular interest to us are the 1st and 99th quantiles of each
distributional statistic, as these are likely to be the most severe
violations of any marginal assumptions that we may make. Figure~\ref{fig:9}
also validates our use of the Student's-t distribution to capture
excess kurtosis in the marginals and the Skew-Student's-t distribution
to capture excess skew.

\subsection{Model Fitting Performance\label{sub:Model-Fitting-Performance}}

To demonstrate that the more generalised structure of the CDCV model
is capable of outperforming the CAVA model's rigid hierarchy, we first
replicate here the primary measures of performance analysis that \cite{Heinen2008}
employed, before extending the analysis to consider other aspects
of model performance.

\begin{table}[H]
\centering
\textcolor{black}{\scriptsize{}}%
\begin{tabular*}{35pc}{@{\extracolsep{\fill}}ll|cccccccc}
 & \multicolumn{1}{l}{} & \multicolumn{8}{c}{\textcolor{black}{\scriptsize{Distribution of Bivariate Rank Correlations}}} \\
\hline 
\textcolor{black}{\scriptsize{Conditioning}} & \textcolor{black}{\scriptsize{Model}} &  & \textcolor{black}{\scriptsize{Mean}} & \textcolor{black}{\scriptsize{Std Dev}} & \textcolor{black}{\scriptsize{q1}} & \textcolor{black}{\scriptsize{q25}} & \textcolor{black}{\scriptsize{q50}} & \textcolor{black}{\scriptsize{q75}} & \textcolor{black}{\scriptsize{q99}} \\
\hline 
\hline 
{\scriptsize{None}} & {\scriptsize{Both}} &  & {\scriptsize{0.3506}} & {\scriptsize{0.1258}} & {\scriptsize{-0.0246}} & {\scriptsize{0.2671}} & {\scriptsize{0.3474}} & {\scriptsize{0.4291}} & {\scriptsize{0.8597}} \\
\hline 
\multirow{2}{*}{{\scriptsize{Market}}} & {\scriptsize{CDCV}} &  & {\scriptsize{0.0022}} & {\scriptsize{0.1575}} & {\scriptsize{-0.4387}} & {\scriptsize{-0.0971}} & {\scriptsize{-0.0037}} & {\scriptsize{0.0860}} & {\scriptsize{0.7638}} \\
 & {\scriptsize{CAVA}} &  & {\scriptsize{0.0116}} & {\scriptsize{0.1505}} & {\scriptsize{-0.4060}} & {\scriptsize{-0.0814}} & {\scriptsize{0.0009}} & {\scriptsize{0.0841}} & {\scriptsize{0.7782}} \\
\hline 
{\scriptsize{Market + Cluster}} & \textbf{\scriptsize{CDCV}} &  & \textbf{\scriptsize{-0.0023}} & \textbf{\scriptsize{0.0936}} & \textbf{\scriptsize{-0.4319}} & \textbf{\scriptsize{-0.0625}} & \textbf{\scriptsize{-0.0016}} & \textbf{\scriptsize{0.0588}} & \textbf{\scriptsize{0.3795}} \\
{\scriptsize{Market + Sector}} & {\scriptsize{CAVA}} &  & {\scriptsize{0.0013}} & {\scriptsize{0.0950}} & {\scriptsize{-0.4835}} & {\scriptsize{-0.0606}} & {\scriptsize{0.0020}} & {\scriptsize{0.0642}} & {\scriptsize{0.3657}} \\
\hline 
 & \multicolumn{1}{l}{} & \multicolumn{8}{c}{\textcolor{black}{\scriptsize{Distribution of (Absolute) Bivariate
Rank Correlations}}} \\
\hline 
\textcolor{black}{\scriptsize{Conditioning}} & \textcolor{black}{\scriptsize{Model}} &  & \textcolor{black}{\scriptsize{Mean}} & \textcolor{black}{\scriptsize{Std Dev}} & \textcolor{black}{\scriptsize{q1}} & \textcolor{black}{\scriptsize{q25}} & \textcolor{black}{\scriptsize{q50}} & \textcolor{black}{\scriptsize{q75}} & \textcolor{black}{\scriptsize{q99}} \\
\hline 
\hline 
{\scriptsize{Market + Cluster}} & \textbf{\scriptsize{CDCV}} &  & \textbf{\scriptsize{0.0733}} & \textbf{\scriptsize{0.0583}} & \textbf{\scriptsize{0.0001}} & \textbf{\scriptsize{0.0287}} & \textbf{\scriptsize{0.0607}} & \textbf{\scriptsize{0.1040}} & \textbf{\scriptsize{0.4670}} \\
{\scriptsize{Market + Sector}} & {\scriptsize{CAVA}} &  & {\scriptsize{0.0747}} & {\scriptsize{0.0587}} & {\scriptsize{0.0001}} & {\scriptsize{0.0296}} & {\scriptsize{0.0625}} & {\scriptsize{0.1065}} & {\scriptsize{0.4981}} \\
\hline 
\end{tabular*}

%\bigskip{}
%\medskip{}

\caption{{Distributional statistics of all bivariate asset correlations,
at various stages of the conditioning process employed when fitting
CAVA and CDCV (clusters$=15$, noise parameter $\Upsilon=11$) models to a rolling learning period of 150 days.}}
\label{tab:2}
\end{table}

\begin{figure}[H]
\centering
\textbf{\includegraphics[scale=0.45]{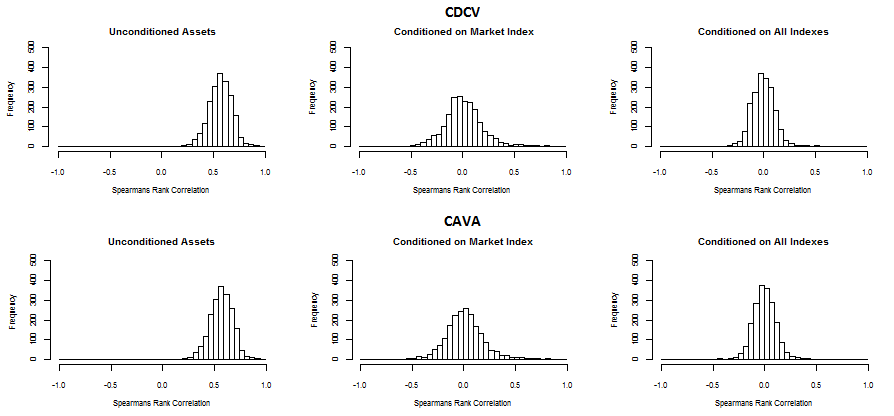}}

\caption{Distributions of all bivariate Spearman's Rho rank
correlations between assets, at various stages of the conditioning
process employed when fitting the CAVA and CDCV models
to a rolling learning period of 150 days.}
\label{fig:10}
\end{figure}
In each time step of our data, we fit both the CDCV and CAVA implementations
to a rolling learning period of $150$ daily returns. The CDCV model
parameters are chosen to include a Kendall's Tau based distance metric,
an Adapted Single Linkage Criterion, a fixed number of clusters set
at $15$ and a volatility-weighted mean index construction with a
noise parameter of $\Upsilon=1/\lambda=11$. These parameters were
chosen based on a cursory performance analysis and will be used throughout
this section before we analyse optimal parameter choices in Section~\ref{sub:Sensitivity-Analysis}. 
We then record the distribution of
bivariate Spearman's Rho rank correlations remaining between pairs
of asset return time series after conditioning our data on first the
market index and then the sector/cluster indexes as illustrated in
Figure~\ref{fig:10}. The resulting distributive statistics are
then summarised across all $850$ time steps in Table~\ref{tab:2}, as an indicator
of the model's ability to capture the dependence in our data set.\\
\\
These results indicate that while the market index conditioning of
the CDCV implementation is out-performed slightly by the CAVA implementation,
the fully conditioned results of the CDCV improve upon those of the
CAVA implementation despite having only performed a cursory parameter
analysis. In particular, the CDCV implementation results in a slightly
lower standard deviation of $0.0936$ as opposed to the CAVA's $0.0950$.
When the absolute rank correlations are considered, the mean, standard
deviation and all quantile values are lower than the corresponding
CAVA results, with the remaining maximum absolute correlation $6.2\%$
lower than the equivalent CAVA value. The CDCV model's absolute q50
percentile value represents a $2.8\%$ drop in the bivariate correlation
remaining, while the absolute q25 percentile value represents a $3.0\%$
drop. The graphical summary of these results, presented in Figure~\ref{fig:10},
illustrates that the CDCV and CAVA implementations also lead to similar
distributions of remaining bivariate correlations. However, in order
to assess more fully the performance of the two models we must analyse
how these distributions vary through time.

\subsection{Stability Analysis\label{sub:Stability-Analysis}}

\begin{figure}[H]
\centering
\textbf{\includegraphics[scale=0.45]{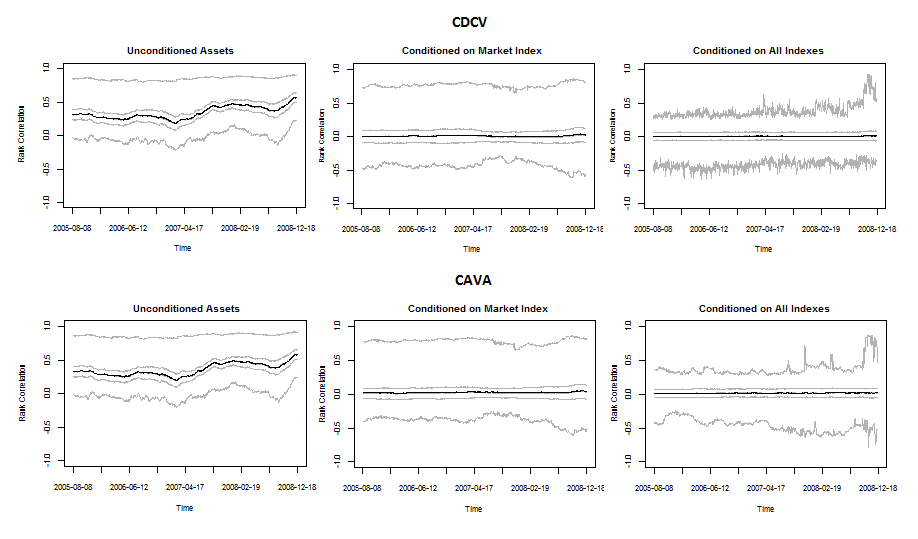}}
\caption{{Evolution through time of the distributional statistics
(q1, q25, q50, q75, q99) of all bivariate correlations between
assets, at various stages of the conditioning process employed when
fitting the CAVA and heuristically optimised CDCV models
to a rolling learning period of 150 days.}}
\label{fig:11}
\end{figure}

In Figure~\ref{fig:11}, we show the evolution of the CDCV bivariate
rank correlation quantile values from Table~\ref{tab:2}, illustrating how the
range of rank correlations in the data varies through time.

In line with Figure~\ref{fig:10}, the conditioning on the market
index in Figure~\ref{fig:11} appears to consistently shift the
mean and median of the conditioned distribution towards zero, while
introducing a positive skew in the correlation distribution. The subsequent
conditioning on the cluster indexes then significantly reduces the
skew, while focusing the 25th and 75th quantiles more closely around
zero. While the model fitting looks largely stable, some minor variability
is introduced in the quantiles due to the flexibility of the CDCV
model's structure, which is allowed to vary in each time step. However,
this analysis also indicates that the CDCV model produces more stable
values than the CAVA structure, which may be due to the dampening effect of 
the noise term used when constructing the CDCV model's derived indexes 
(see Section \ref{sub:Deriving-Hierarchical-Indexes}).
Figure~\ref{fig:11} also illustrates that both our implementations
are equally unable to capture the most extreme positive correlations
that occur in late 2008. As our analysis is primarily comparative
we do not address this point further here, but an area for further
research would be to investigate further whether the model could be
improved to also capture these most extreme dependencies, for example
by selecting from a larger set of bivariate copulas.

\begin{table}[H]
\begin{centering}
\textcolor{black}{\scriptsize{}}%
\begin{tabular*}{35pc}{@{\extracolsep{\fill}}ll|ccccccccc|c}
\hline 
 & \textcolor{black}{\tiny{Model}} & \textcolor{black}{\tiny{8-8-05}} & \textcolor{black}{\tiny{9-1-06}} & \textcolor{black}{\tiny{12-6-06}} & \textcolor{black}{\tiny{9-11-06}} & \textcolor{black}{\tiny{17-4-07}} & \textcolor{black}{\tiny{17-9-07}} & \textcolor{black}{\tiny{19-2-08}} & \textcolor{black}{\tiny{21-7-08}} & \textcolor{black}{\tiny{18-12-08}} & \textcolor{black}{\tiny{Mean}} \\
\hline 
\hline 
\multirow{1}{*}{} & {\tiny{CDCV}} & {\tiny{166}} & {\tiny{177}} & {\tiny{166}} & {\tiny{168}} & {\tiny{186}} & {\tiny{190}} & {\tiny{179}} & {\tiny{192}} & {\tiny{215}} & {\tiny{182}} \\
\multirow{1}{*}{} & {\tiny{CAVA}} & {\tiny{157}} & {\tiny{172}} & {\tiny{172}} & {\tiny{172}} & {\tiny{181}} & {\tiny{177}} & {\tiny{176}} & {\tiny{185}} & {\tiny{220}} & {\tiny{157}} \\
\hline 
\end{tabular*}
\par\end{centering}{\scriptsize \par}

\caption{The number of copula parameters fitted
by the heuristically optimised (15 clusters; noise
parameter, $\Upsilon=11$) CDCV and the CAVA models through time.}
\label{tab:3}
\end{table}

A final component of this analysis is detailed in Table~\ref{tab:3}, which illustrates
the variation in the number of parameters to be fitted through time.
The number of parameters utilised by both the CDCV and CAVA implementations
increases substantially during times of market stress, primarily due
to the increase in the number of Student's-t copulas selected.

\subsection{VaR Backtesting\label{sub:VaR-Backtesting}}

Another comparison that we provide between the CDCV and CAVA implementations
is their Value-at-Risk (VaR) backtesting performance, based on an
analysis of the number of VaR breaches that occur during within-sample
and out-of-sample testing and the associated Proportion of Failures
(PoF) test statistic of unconditional coverage given by \cite{Kupiec1995}
as

\[
LR_{PoF}=-2\ln\left(\frac{\left(1-q\right)^{T-x}q^{x}}{\left(1-\left(\frac{x}{T}\right)\right)^{T-x}\left(\frac{x}{T}\right)^{x}}\right)\,,
\]
where $x$ is the number of exceptions, $T$ is the total number of
trials (time steps) and the test statistic is asymptotically distributed
as $LR_{PoF}\thicksim\chi_{1df}^{2}$. As our analysis is focused
on the performance of high-dimensional portfolios we are content for
now to calculate the vector of theoretical VaR quantiles using an
equally weighted portfolio of all 62 assets considered in this analysis.
We leave a more thorough review of sub-portfolio backtesting performance
and conditional coverage as topics for further analysis.\\
\\
When testing the CDCV model within-sample, we obtain a $VaR_{95}$
p-value of $0.250$ under the null hypothesis $H_{0}$ that the actual
exception rate $q$ equals the observed exceptions rate $\hat{q}$
(where an exception is deemed to be a breach of the predicted $95$\%
VaR threshold), and thus we can comfortably accept $H_{0}$ at any
reasonable level of confidence (see Table~\ref{tab:4}). When considering the
more extreme $99^{th}$ percentile loss we obtain a $VaR_{99}$ p-value
of $0.043$ and so would narrowly reject $H_{0}$ at the $95\%$ confidence
level, while continuing to accept it if we test at the $99\%$ confidence
level.

\begin{figure}[H]
\centering
\includegraphics[scale=0.58]{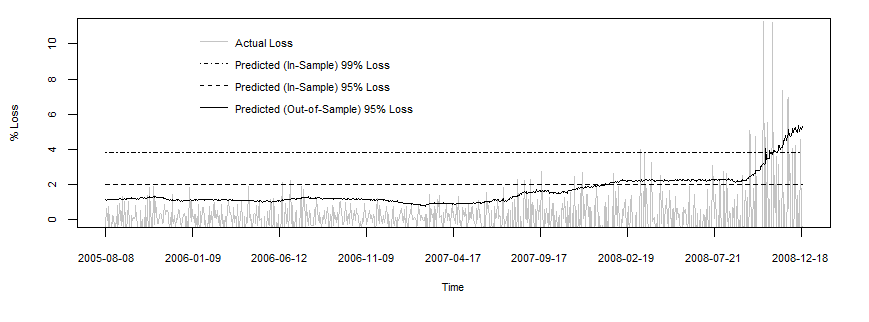}

\caption{{VaR back-testing performance of the CDCV
model: Actual \% losses plotted against the
out-of-sample 95th percentile loss vector plus the within-sample predicted
95th and 99th percentile loss vectors, based on an equally weighted
portfolio of all 62 in-scope assets.}}
\label{fig:12}
\end{figure}

\begin{table}[H]
\begin{centering}
\textcolor{black}{\scriptsize{}}%
\begin{tabular*}{35pc}{@{\extracolsep{\fill}}ll|c|cccccccc}
\hline 
 & \textcolor{black}{\scriptsize{Model}} & \textcolor{black}{\scriptsize{$\alpha$}} &  & \textcolor{black}{\scriptsize{$VaR_{\alpha}$}} & \textcolor{black}{\scriptsize{Hits}} & \textcolor{black}{\scriptsize{Hit \%}} & {\scriptsize{$LR_{POF}$}} & {\scriptsize{p-Value}} & \textcolor{black}{\scriptsize{95\% Conf}} & \textcolor{black}{\scriptsize{99\% Conf}} \\
\hline 
\hline 
 & \textbf{\textcolor{black}{\scriptsize{CDCV}}} & \textbf{\textcolor{black}{\scriptsize{95}}} &  & \textbf{\textcolor{black}{\scriptsize{1.97}}} & \textbf{\textcolor{black}{\scriptsize{50}}} & \textbf{\textcolor{black}{\scriptsize{5.88}}} & \textbf{\scriptsize{1.322}} & \textbf{\scriptsize{0.250}} & \textbf{\textcolor{black}{\scriptsize{Accept $H_{0}$}}} & \textbf{\textcolor{black}{\scriptsize{Accept $H_{0}$}}} \\
 & \textcolor{black}{\scriptsize{CAVA}} & \textcolor{black}{\scriptsize{95}} &  & {\scriptsize{1.90}} & {\scriptsize{52}} & {\scriptsize{6.12}} & {\scriptsize{2.093}} & {\scriptsize{0.147}} & \textcolor{black}{\scriptsize{Accept $H_{0}$}} & \textcolor{black}{\scriptsize{Accept $H_{0}$}} \\
\hline 
 & \textbf{\textcolor{black}{\scriptsize{CDCV}}} & \textbf{\textcolor{black}{\scriptsize{99}}} &  & \textbf{\textcolor{black}{\scriptsize{3.80}}} & \textbf{\textcolor{black}{\scriptsize{15}}} & \textbf{\textcolor{black}{\scriptsize{1.76}}} & \textbf{\textcolor{black}{\scriptsize{4.090}}} & \textbf{\textcolor{black}{\scriptsize{0.043}}} & \textbf{\textcolor{black}{\scriptsize{Reject $H_{0}$}}} & \textbf{\textcolor{black}{\scriptsize{Accept $H_{0}$}}} \\
 & \textcolor{black}{\scriptsize{CAVA}} & \textcolor{black}{\scriptsize{99}} &  & \textcolor{black}{\scriptsize{3.70}} & \textcolor{black}{\scriptsize{16}} & \textcolor{black}{\scriptsize{1.88}} & \textcolor{black}{\scriptsize{5.308}} & \textcolor{black}{\scriptsize{0.021}} & \textcolor{black}{\scriptsize{Reject $H_{0}$}} & \textcolor{black}{\scriptsize{Accept $H_{0}$}} \\
\hline 
\end{tabular*}
\par\end{centering}{\scriptsize \par}

\caption{Within-sample Value at Risk ($VaR_{95}$
\& $VaR_{99}$) backtesting results for the CDCV and CAVA implementations.
``Hits'' are deemed to be breaches of the relevant $VaR_{\alpha}$
threshold.}
\label{tab:4}
\end{table}

In the context of this analysis, we may equate the acceptance of $H_{0}$
to a validation of the $VaR_{\alpha}$ number(s) generated by the
model, indicating that the model fits the historical data sufficiently,
in so far as that can be assessed by considering the $\alpha-level$
percentile loss. Repeating this test for our model using the CAVA
industry hierarchy, we also accept $H_{0}$ under the same conditions
as for the CDCV model, albeit with the lower $VaR_{95}$ and $VaR_{99}$
p-values of $0.147$ and $0.021$ respectively. This suggests that
the CDCV implementation provides a slightly better PoF backtesting
performance than the CAVA implementation for this set of test data.
\\
\\
When we extend this analysis to consider out-of-sample testing by
fitting a rolling $150$-day window and generating the predicted $VaR_{95}$
value in each time step, we obtain a breach percentage of $8.23$\%
($70$ breaches) for both the CDCV and the CAVA models, with a resultant
Kupiec test statistic of $LR_{POF}=15.806$ and a p-value of $\leq0.01$,
leading us to clearly reject the null hypothesis. This is reflective
of the difficulties of predictive modelling, and the effects of model
risk during significant market shifts or downturns, as illustrated
in Figure~\ref{fig:12}. If we consider the first $750$ time
steps only (and disregard the final $100$ which represent the beginning
of the 2008 financial crisis), we obtain an out-of-sample $VaR_{95}$
breach percentage of $6.8$\%, which gives a Kupiec test statistic
of $LR_{POF}=4.621$ and results in a p-value of $0.032$. Under these
circumstances, we would accept $H_{0}$ at the $99$\% confidence
level, but continue to reject it (and accept the alternative hypothesis
$H_{1}$) at the $95$\% confidence level.

\subsection{Copula Fitting \& Selection Analysis\label{sub:Copula-Fitting-&}}

We next extend our analysis of the CDCV and CAVA model fitting evolution
by reviewing the variation through time in the percentage of copulas
selected for both the unconditional copulas in the first tree and
the conditional copulas in the subsequent trees.

\begin{figure}[H]
\centering
\includegraphics[scale=0.52]{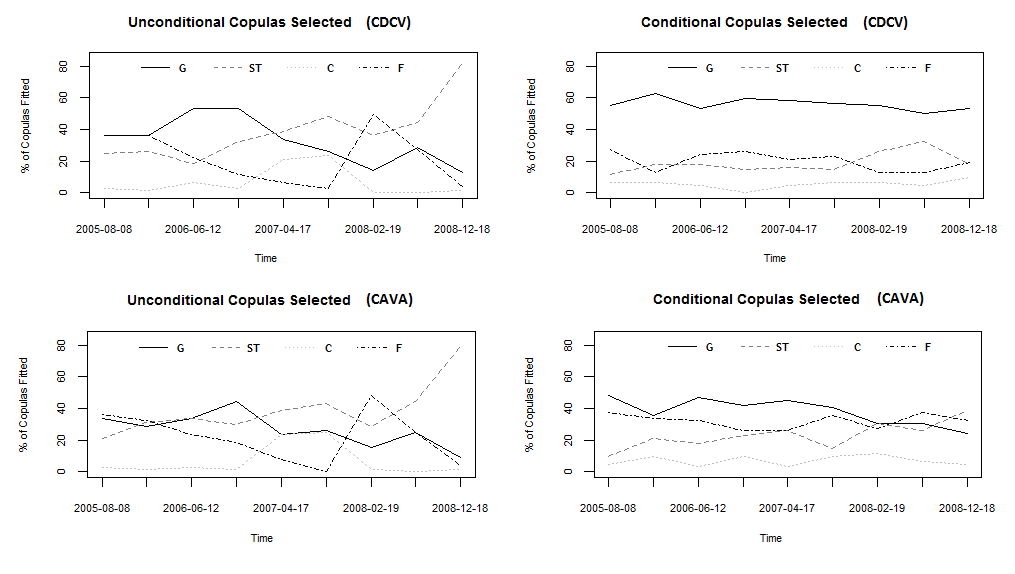}
\caption{Evolution through time of the number
of unconditional and conditional copulas fitted as Gaussian (G), Student's-t
(ST), Clayton (C) or Frank (F) by the CDCV and CAVA models, where
we have discretised the data into 9 time steps between 8th August
2005 and the 18th December 2008 in order to minimise noise and clearly
illustrate trends.}
\label{fig:13}
\end{figure}

As shown in Table~\ref{tab:5} and Figure~\ref{fig:13}, the primary difference
between the CDCV and CAVA copula fitting evolutions is that after
conditioning on the market index, the CDCV model then selects a largely
time-consistent proportion of each copula type, with Gaussian copulas
being selected $56$\% of the time. In comparison, the CAVA implementation
selects Gaussian copulas only $39$\% of the time (on average), and
the actual number selected decreases over $20$ percentage points
throughout the backtesting time frame, while the number of Student's-t
copulas increases by over $20$ percentage points. This may be attributable
to the direct link between the CDCV model's derived cluster indexes
and the market index, which is a weighted average of the cluster indexes.
As this similarity can reasonably assumed to be more pronounced than
between the S\&P 500 and S\&P Industry Sector indexes (due to the inclusion
of many other equities in those indexes), we may expect that the first
level of conditioning on the CDCV model's market index would capture
a greater proportion of the non-Gaussian cluster index behaviours
than in the CAVA framework. 

\begin{table}[H]
\begin{centering}
\textcolor{black}{\scriptsize{}}%
\begin{tabular*}{35pc}{@{\extracolsep{\fill}}llll|ccccc}
\hline 
 & \textcolor{black}{\scriptsize{Model}} & \textcolor{black}{\scriptsize{Root Node}} &  & \textcolor{black}{\scriptsize{G (\%)}} & \textcolor{black}{\scriptsize{ST (\%)}} & \textcolor{black}{\scriptsize{C (\%)}} & \textcolor{black}{\scriptsize{F (\%)}} &  \\
\hline 
\multirow{2}{*}{} & \multirow{2}{*}{\textcolor{black}{\scriptsize{CDCV}}} & \textcolor{black}{\scriptsize{Market Index}} &  & \textcolor{black}{\scriptsize{35.58}} & \textcolor{black}{\scriptsize{36.51}} & \textcolor{black}{\scriptsize{6.21}} & \textcolor{black}{\scriptsize{21.70}} &  \\
 &  & \textcolor{black}{\scriptsize{Cluster Indexes}} &  & \textcolor{black}{\scriptsize{56.09}} & \textcolor{black}{\scriptsize{19.06}} & \textcolor{black}{\scriptsize{5.47}} & \textcolor{black}{\scriptsize{19.37}} &  \\
\hline 
\multirow{2}{*}{} & \multirow{2}{*}{\textcolor{black}{\scriptsize{CAVA}}} & \textcolor{black}{\scriptsize{Market Index}} &  & \textcolor{black}{\scriptsize{30.49}} & \textcolor{black}{\scriptsize{39.78}} & \textcolor{black}{\scriptsize{6.81}} & \textcolor{black}{\scriptsize{22.92}} &  \\
 &  & \textcolor{black}{\scriptsize{Cluster Indexes}} &  & \textcolor{black}{\scriptsize{39.02}} & \textcolor{black}{\scriptsize{23.61}} & \textcolor{black}{\scriptsize{8.39}} & \textcolor{black}{\scriptsize{28.98}} &  \\
\hline 
\end{tabular*}
\par\end{centering}{\scriptsize \par}

\caption{Percentages (averaged across all time
steps) of unconditional and conditional copulas fitted as Gaussian
(G), Student's-t (ST), Clayton (C) or Frank (F) by the CDCV and CAVA
models in a given time step, based on a 150 day rolling learn period.}
\label{tab:5}
\end{table}

Another notable feature of the unconditional copulas selected by both
models over the sample period is that the number of Student's-t copulas
fitted increases roughly in response to the increase in market turbulence.
At the height of the financial crisis in 2008, the {Student's-t} copula
accounted for 66 (86\%) of the CDCV model's unconditional copulas
between the market index and both asset returns and cluster indexes,
as shown in Table 6. 

\begin{table}[H]
\begin{centering}
\textcolor{black}{\scriptsize{}}%
\begin{tabular*}{35pc}{@{\extracolsep{\fill}}ll|ccccccccc|c}
\hline 
\textcolor{black}{\tiny{Root Node}} & \textcolor{black}{\tiny{Copula}} & \textcolor{black}{\tiny{8-8-05}} & \textcolor{black}{\tiny{9-1-06}} & \textcolor{black}{\tiny{12-6-06}} & \textcolor{black}{\tiny{9-11-06}} & \textcolor{black}{\tiny{17-4-07}} & \textcolor{black}{\tiny{17-9-07}} & \textcolor{black}{\tiny{19-2-08}} & \textcolor{black}{\tiny{21-7-08}} & \textcolor{black}{\tiny{18-12-08}} & \textcolor{black}{\tiny{Mean}} \\
\hline 
\hline 
\multirow{4}{*}{\textcolor{black}{\tiny{Market Index}}} & \textcolor{black}{\tiny{G}} & \textcolor{black}{\tiny{27}} & \textcolor{black}{\tiny{24}} & \textcolor{black}{\tiny{42}} & \textcolor{black}{\tiny{47}} & \textcolor{black}{\tiny{25}} & \textcolor{black}{\tiny{19}} & \textcolor{black}{\tiny{12}} & \textcolor{black}{\tiny{21}} & \textcolor{black}{\tiny{7}} & {\tiny{21.70}} \\
 & \textcolor{black}{\tiny{ST}} & \textcolor{black}{\tiny{21}} & \textcolor{black}{\tiny{21}} & \textcolor{black}{\tiny{14}} & \textcolor{black}{\tiny{19}} & \textcolor{black}{\tiny{32}} & \textcolor{black}{\tiny{38}} & \textcolor{black}{\tiny{28}} & \textcolor{black}{\tiny{36}} & \textcolor{black}{\tiny{66}} & \textcolor{black}{\tiny{27.40}} \\
 & \textcolor{black}{\tiny{C}} & \textcolor{black}{\tiny{2}} & \textcolor{black}{\tiny{1}} & \textcolor{black}{\tiny{4}} & \textcolor{black}{\tiny{1}} & \textcolor{black}{\tiny{14}} & \textcolor{black}{\tiny{18}} & \textcolor{black}{\tiny{0}} & \textcolor{black}{\tiny{0}} & \textcolor{black}{\tiny{1}} & \textcolor{black}{\tiny{4.78}} \\
 & \textcolor{black}{\tiny{F}} & \textcolor{black}{\tiny{27}} & \textcolor{black}{\tiny{31}} & \textcolor{black}{\tiny{17}} & \textcolor{black}{\tiny{10}} & \textcolor{black}{\tiny{6}} & \textcolor{black}{\tiny{2}} & \textcolor{black}{\tiny{37}} & \textcolor{black}{\tiny{20}} & \textcolor{black}{\tiny{3}} & \textcolor{black}{\tiny{16.71}} \\
\hline 
\multirow{4}{*}{\textcolor{black}{\tiny{Cluster Indexes}}} & \textcolor{black}{\tiny{G}} & \textcolor{black}{\tiny{41}} & \textcolor{black}{\tiny{33}} & \textcolor{black}{\tiny{28}} & \textcolor{black}{\tiny{33}} & \textcolor{black}{\tiny{34}} & \textcolor{black}{\tiny{37}} & \textcolor{black}{\tiny{38}} & \textcolor{black}{\tiny{33}} & \textcolor{black}{\tiny{36}} & \textcolor{black}{\tiny{34.78}} \\
 & \textcolor{black}{\tiny{ST}} & \textcolor{black}{\tiny{5}} & \textcolor{black}{\tiny{16}} & \textcolor{black}{\tiny{12}} & \textcolor{black}{\tiny{9}} & \textcolor{black}{\tiny{14}} & \textcolor{black}{\tiny{12}} & \textcolor{black}{\tiny{11}} & \textcolor{black}{\tiny{16}} & \textcolor{black}{\tiny{9}} & \textcolor{black}{\tiny{11.82}} \\
 & \textcolor{black}{\tiny{C}} & \textcolor{black}{\tiny{4}} & \textcolor{black}{\tiny{2}} & \textcolor{black}{\tiny{4}} & \textcolor{black}{\tiny{1}} & \textcolor{black}{\tiny{3}} & \textcolor{black}{\tiny{4}} & \textcolor{black}{\tiny{3}} & \textcolor{black}{\tiny{4}} & \textcolor{black}{\tiny{5}} & \textcolor{black}{\tiny{3.39}} \\
 & \textcolor{black}{\tiny{F}} & \textcolor{black}{\tiny{12}} & \textcolor{black}{\tiny{11}} & \textcolor{black}{\tiny{18}} & \textcolor{black}{\tiny{19}} & \textcolor{black}{\tiny{11}} & \textcolor{black}{\tiny{9}} & \textcolor{black}{\tiny{10}} & \textcolor{black}{\tiny{9}} & \textcolor{black}{\tiny{12}} & \textcolor{black}{\tiny{12.01}} \\
\hline 
\multirow{2}{*}{\textcolor{black}{\tiny{Joint-Simplified}}} & \textcolor{black}{\tiny{G}} & \textcolor{black}{\tiny{0}} & \textcolor{black}{\tiny{0}} & \textcolor{black}{\tiny{0}} & \textcolor{black}{\tiny{0}} & \textcolor{black}{\tiny{0}} & \textcolor{black}{\tiny{0}} & \textcolor{black}{\tiny{0}} & \textcolor{black}{\tiny{0}} & \textcolor{black}{\tiny{0}} & \textcolor{black}{\tiny{0}} \\
 & \textcolor{black}{\tiny{ST}} & \textcolor{black}{\tiny{1}} & \textcolor{black}{\tiny{1}} & \textcolor{black}{\tiny{1}} & \textcolor{black}{\tiny{1}} & \textcolor{black}{\tiny{1}} & \textcolor{black}{\tiny{1}} & \textcolor{black}{\tiny{1}} & \textcolor{black}{\tiny{1}} & \textcolor{black}{\tiny{1}} & \textcolor{black}{\tiny{1}} \\
\hline 
\end{tabular*}
\par\end{centering}{\scriptsize \par}

\caption{Numbers of bivariate copulas fitted
to the previous 150 business days (the learning period) for each of
the 9 discretised time steps illustrated earlier in Figure~\ref{fig:13},
broken out into copulas rooted at the market node, copulas rooted
at the cluster nodes and multivariate jointly-simplifying copulas.}
\end{table}

In our analysis we also see interesting swells in the selection of
first Clayton and then Frank copulas in 2007 and then early 2008 respectively,
where the increase in the numbers of Clayton copulas coincides with
the maximal negative skew and kurtosis values observed in the marginal
data, and the increase in the number of Frank copulas selected coincides
with the steady decrease in skew and kurtosis illustrated earlier in Figure~\ref{fig:9}. 
This behaviour seems reasonable, as we would expect Clayton copulas to
be selected when there is increased negative tail dependence between
both assets and indexes, and the elliptical Frank copula to be selected
when such characteristics are reduced. Table 6 also indicates that,
for our data set, the Student's-t multivariate copula is almost always
more appropriate than the Gaussian copula for joint simplification.
This is reflected in the AIC scores from which the model fitting selection
is derived and is consistent with the analysis of \cite{Brechmann2013a}
which highlights that the choice of a joint Gaussian simplification
(per \cite{Heinen2008}) is often not appropriate.

\subsection{Marginal Fitting \& Selection Analysis\label{sub:Marginal-Fitting-&}}

While the analysis of the marginal distributions over all time steps
in Figure~\ref{fig:8} indicates that we will rarely need to
fit Gaussian marginal distributions, we must recall that (as indicated
by Figure~\ref{fig:9}) the distributions will vary through time
when fitted to a rolling learning period.

\begin{figure}[H]
\centering
\includegraphics[scale=0.47]{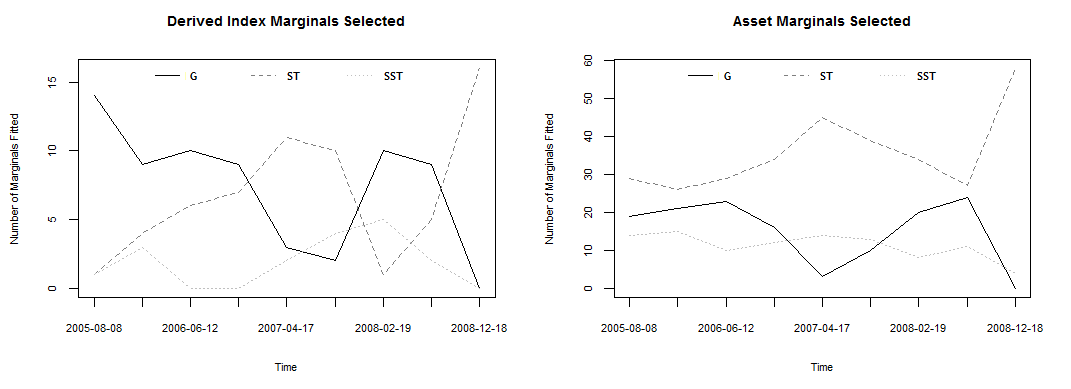}
\caption{Number of index and asset marginal
distributions fitted as Gaussian (G), Student's-t (ST) and Skew {Student's-t}
(SST) by the CDCV model, where we have discretised the data into 9
time steps between 8th August 2005 and the 18th December 2008 in order
to minimise noise and clearly illustrate trends.}
\label{fig:14}
\end{figure}

In Figure~\ref{fig:14} we have again discretised the data into
9 time steps between \textcolor{black}{8th August 2005 and the 18th
December 2008 in order to minimise noise and clearly illustrate the
trends in the number of Gaussian (G), Student's-t (ST) and Skew-Student's-t
(SST) marginal distributions selected by the CDCV model via AIC. As
we might expect, the number of Gaussian marginals selected tends to
decrease overall, and does so sharply in 2008 as the financial crisis
took hold and non-Gaussian characteristics became more prevalent in
the asset returns. The number of Student's-t marginals selected is
shown in Figure~\ref{fig:14} to approximately mirror the number
of Gaussian marginals selected in that one increases when the other
decreases. In the case of the derived cluster indexes (constructed
via a volatility-weighted averaging of their constituent cluster assets),
we also see similar selection patterns to the asset marginals themselves,
but with a relatively increased proportion of Gaussian distributions.}

\begin{table}[H]
\begin{centering}
\textcolor{black}{\scriptsize{}}%
\begin{tabular*}{35pc}{@{\extracolsep{\fill}}ll|ccccccccc|c}
\hline 
\textcolor{black}{\tiny{Marginal}} & \textcolor{black}{\tiny{Distrib}} & \textcolor{black}{\tiny{8-8-05}} & \textcolor{black}{\tiny{9-1-06}} & \textcolor{black}{\tiny{12-6-06}} & \textcolor{black}{\tiny{9-11-06}} & \textcolor{black}{\tiny{17-4-07}} & \textcolor{black}{\tiny{17-9-07}} & \textcolor{black}{\tiny{19-2-08}} & \textcolor{black}{\tiny{21-7-08}} & \textcolor{black}{\tiny{18-12-08}} & \textcolor{black}{\tiny{Mean}} \\
\hline 
\hline 
\multirow{3}{*}{\textcolor{black}{\tiny{Assets (62)}}} & \textcolor{black}{\tiny{G}} & \textcolor{black}{\tiny{19}} & \textcolor{black}{\tiny{21}} & \textcolor{black}{\tiny{23}} & \textcolor{black}{\tiny{16}} & \textcolor{black}{\tiny{3}} & \textcolor{black}{\tiny{10}} & \textcolor{black}{\tiny{20}} & \textcolor{black}{\tiny{24}} & \textcolor{black}{\tiny{0}} & \textcolor{black}{\tiny{15.28}} \\
 & \textcolor{black}{\tiny{ST}} & \textcolor{black}{\tiny{29}} & \textcolor{black}{\tiny{26}} & \textcolor{black}{\tiny{29}} & \textcolor{black}{\tiny{34}} & \textcolor{black}{\tiny{45}} & \textcolor{black}{\tiny{39}} & \textcolor{black}{\tiny{34}} & \textcolor{black}{\tiny{27}} & \textcolor{black}{\tiny{58}} & \textcolor{black}{\tiny{34.57}} \\
 & \textcolor{black}{\tiny{SST}} & \textcolor{black}{\tiny{14}} & \textcolor{black}{\tiny{15}} & \textcolor{black}{\tiny{10}} & \textcolor{black}{\tiny{12}} & \textcolor{black}{\tiny{14}} & \textcolor{black}{\tiny{13}} & \textcolor{black}{\tiny{8}} & \textcolor{black}{\tiny{11}} & \textcolor{black}{\tiny{4}} & \textcolor{black}{\tiny{12.57}} \\
\hline 
\multirow{3}{*}{\textcolor{black}{\tiny{Indexes (16)}}} & \textcolor{black}{\tiny{G}} & \textcolor{black}{\tiny{11}} & \textcolor{black}{\tiny{9}} & \textcolor{black}{\tiny{10}} & \textcolor{black}{\tiny{9}} & \textcolor{black}{\tiny{2}} & \textcolor{black}{\tiny{2}} & \textcolor{black}{\tiny{12}} & \textcolor{black}{\tiny{11}} & \textcolor{black}{\tiny{0}} & {\tiny{8.24}} \\
 & \textcolor{black}{\tiny{ST}} & \textcolor{black}{\tiny{2}} & \textcolor{black}{\tiny{7}} & \textcolor{black}{\tiny{6}} & \textcolor{black}{\tiny{6}} & \textcolor{black}{\tiny{11}} & \textcolor{black}{\tiny{9}} & \textcolor{black}{\tiny{2}} & \textcolor{black}{\tiny{3}} & \textcolor{black}{\tiny{16}} & {\tiny{5.25}} \\
 & \textcolor{black}{\tiny{SST}} & \textcolor{black}{\tiny{3}} & \textcolor{black}{\tiny{0}} & \textcolor{black}{\tiny{0}} & \textcolor{black}{\tiny{1}} & \textcolor{black}{\tiny{3}} & \textcolor{black}{\tiny{5}} & \textcolor{black}{\tiny{2}} & \textcolor{black}{\tiny{2}} & \textcolor{black}{\tiny{0}} & {\tiny{2.31}} \\
\hline 
\end{tabular*}
\par\end{centering}{\scriptsize \par}

\caption{Numbers of each marginal distribution
type fitted to the previous 150 business days (the learning period)
for each of the 9 discretised time steps illustrated earlier in Figure~\ref{fig:14},
broken out into marginal distributions of the asset time series and
of the derived index time series.}
\label{tab:8}
\end{table}

\textcolor{black}{Quantification of these marginal distribution selection
results through time is provided in Table~\ref{tab:8}, which suggests that over
all time steps the Student's-t distribution accounts for more than
50\% of all asset marginals selected, while the Gaussian distribution
is selected approximately 50\% of the time when fitting the derived
index marginals. The number of Skew-Student's-t marginals varies less
through time, and is selected for approximately 15\% of the asset
marginals and 20\% of the derived index marginals across all time
steps.}

\subsection{Clustering Analysis\label{sub:Clustering-Analysis}}

While we have left the optimisation of clustering and index construction
methodologies as for further research, we briefly illustrate
here the impact of such approaches on the composition of the clusters obtained.

\begin{figure}[H]
\centering
\includegraphics[scale=0.37]{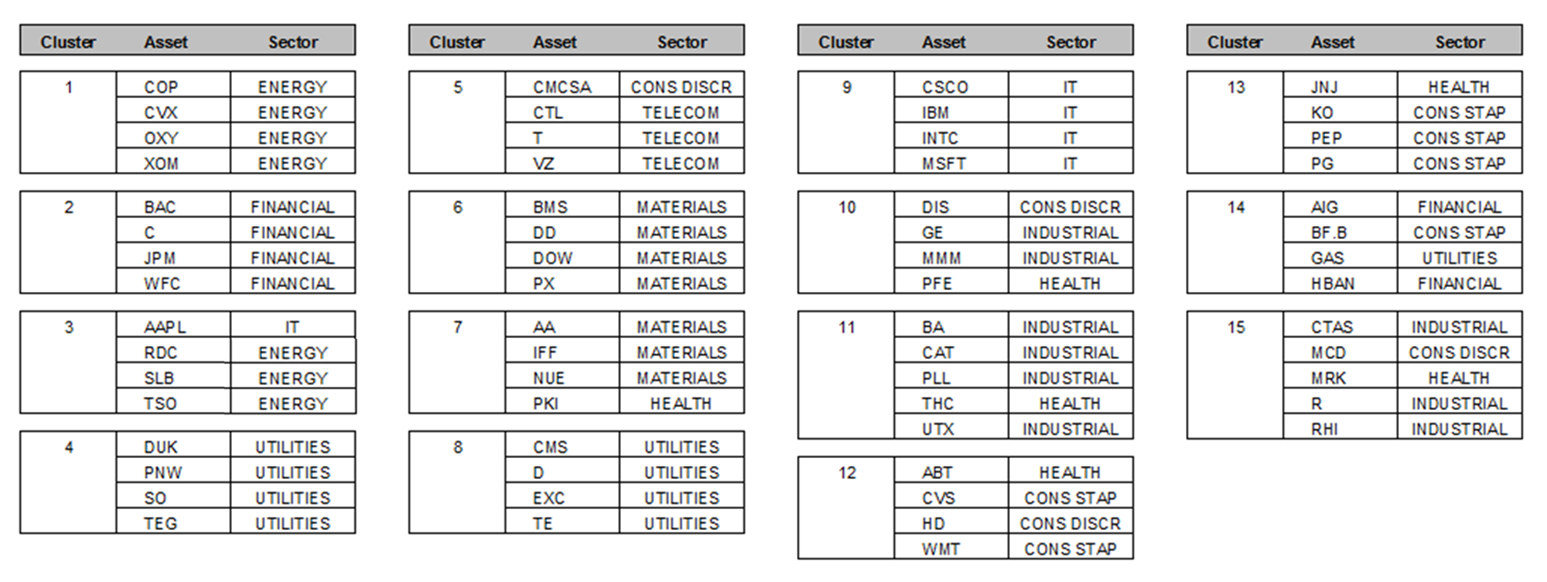}
\caption{The decomposition of 15 clusters generated
using daily returns between 1st January 2005 and
18th December 2008, a Kendall's Tau based distance metric and an Adapted
Single Linkage Criterion.}
\label{fig:15}
\end{figure}

As illustrated in Figure~\ref{fig:15}, the cluster decomposition
obtained when applying the CDCV model to our full analysis time period
includes many clusters that are still constructed from within-industry
assets, i.e. those from the same industry. However, it can be seen that 
many of the assets, particularly Health companies, appear in different 
clusters from their industry peers. We believe that this is a key 
benefit of the CDCV model's clustering approach: within any timestep
clusters may be formed from assets in different industry groups, but
we expect their behaviour to be more closely related during the 
learning period in question.

\subsection{Sensitivity Analysis\label{sub:Sensitivity-Analysis}}

\textcolor{black}{We next analyse the sensitivity of the CDCV model's
performance to changes in its construction. This is an aspect of these
simplified vine copula models that has not been addressed by the existing
literature, despite being fundamental to the usability of such models.
We do not provide an exhaustive analysis here, but rather provide
initial exploratory results that indicate such considerations are
material and may indeed have a significant impact on model performance.}

\begin{figure}[H]
\centering
\includegraphics[scale=0.52]{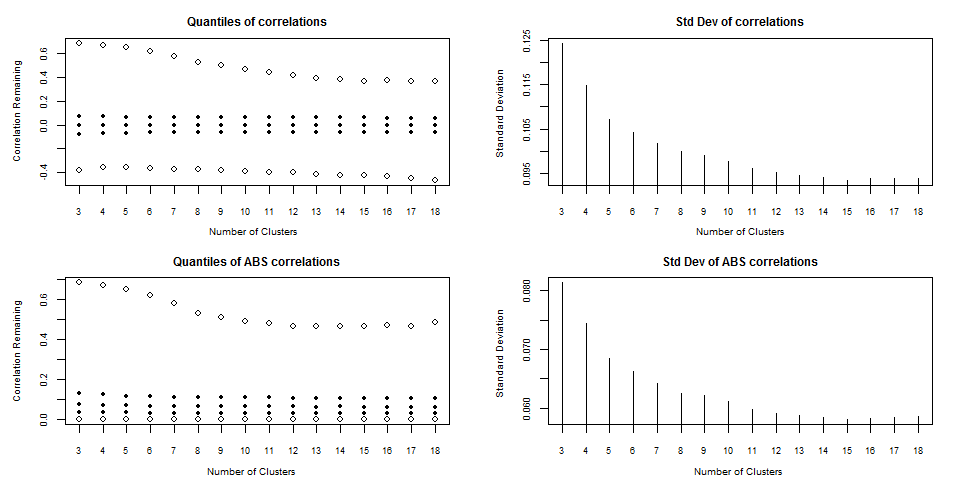}

\caption{For number of clusters $3\leq N\leq18$, we plot the quantiles q1, q25,
q50, q75, q99 and the standard deviation of the distribution of
bivariate rank correlations remaining after conditioning on all nodes
of the CDCV model's vine copula. We also plot equivalent values for
the absolute rank correlations remaining. These summary metrics are
each derived by averaging the equivalent metrics obtained from model
fitting 50 time steps between 1st January 2005 and
18th December 2008, a learning period
of 150 days, a vol-weighted index with a noise parameter of $\Upsilon=11$,
the Adapted Single Linkage Criterion and a Kendall's Tau based distance
metric.}
\label{fig:16}
\end{figure}

\textcolor{black}{In Figure~\ref{fig:16}, we implement the CDCV
model with the same parameter settings as in the analysis above, but
additionally vary the number of clusters that we construct from the
62 asset time series. For ease of implementation we fit each parameter
combination for a sample of 50 time steps and average the results.}
This figure illustrates that the number of clusters utilised for this
data set has a material effect on the standard deviation of the remaining
correlations after model fitting the vine, prior to application of
the jointly simplifying multivariate copula. In the case where only
three clusters were used, this standard deviation rises to 0.12 from
its minimum of 0.093 obtained using 15 clusters, suggesting that model
performance deteriorates as cluster size increases or as the number
of clusters decreases. These results also highlight that when the
minimal cluster size decreases (i.e., the number of clusters increases)
the high negative correlations that we have attempted to minimise
with our index noise parameter(s) are more prevalent. Such findings
indicate that the clustering or grouping of a market-sector vine copula
model has a material impact on model performance and should be considered
by future research in this area.

\begin{figure}[H]
\centering
\includegraphics[scale=0.52]{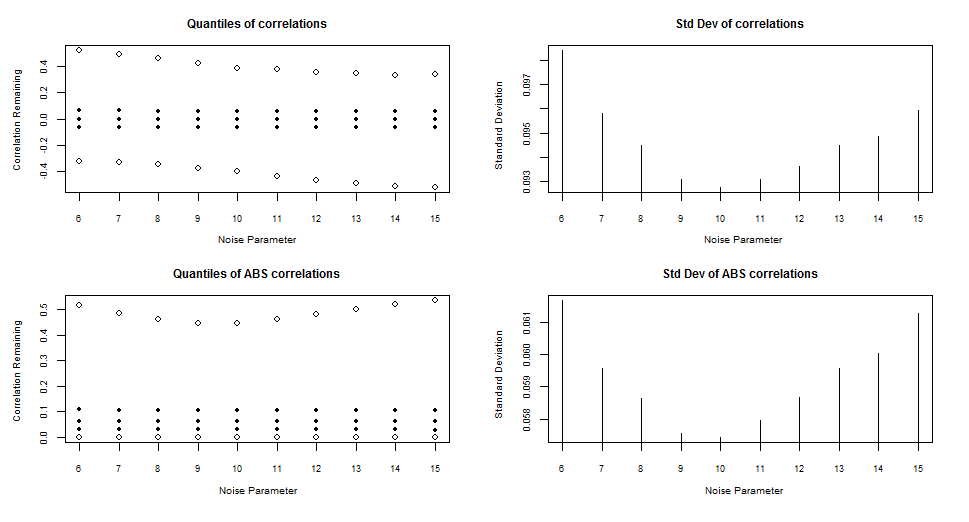}

\caption{For noise parameter $6\leq\Upsilon\leq15$, we plot the quantiles q1, q25,
q50, q75, q99 and the standard deviation of the distribution of
bivariate rank correlations remaining after conditioning on all nodes
of the CDCV model's vine copula. We also plot equivalent values for
the absolute rank correlations remaining. Results are obtained from
the same setup described in Figure~\ref{fig:16}, but with an
additional requirement for 15 clusters.}
\label{fig:17}
\end{figure}

In Figure~\ref{fig:17}, we next vary the scale of the noise
term used to construct our indexes. Varying the noise parameter $\Upsilon=1/\lambda$
defined in (\ref{eq:noise}), we observe that as $\Upsilon\rightarrow\infty$
and thus $I_{+}\rightarrow I$ we are left with high negative correlations
caused by the similarity between asset and index time series. By introducing
increasingly sizable perturbations we suppress the high negative correlations
but we also lose some of the ability to condition away the high positive
correlations in our clusters. Figure~\ref{fig:17} suggests that
the optimal noise parameter range for this data set is $9\leq\Upsilon\leq11$
and illustrates that the detrimental effect of the random noise term
on the standard deviation of remaining bivariate rank correlations
is minimal for the noise parameter range required to dampen the most
extreme negative correlations.

\section{Discussion of Findings and Further Areas for Research\label{sec:Discussion-of-Findings}}

The analysis presented in the preceding section suggests that the
proposed CDCV model is a viable choice for modelling high dimensional
dependence structures in a financial context, and in particular that
it is capable of providing increased flexibility and improved performance
when compared to the original CAVA model of \cite{Heinen2008}. That
such results are obtainable without the use of externally sourced
index time series indicates clearly that future research endeavouring
to implement simplified vine copula models need not and should not
restrict their analysis to specific hierarchical constructions or
data sets. Our analysis demonstrates that the restriction on the clustering
structure imposed by the use of market-available external indexes
has a material impact on the performance of such vine copula models.
In particular, we have shown that the way in which assets are partitioned
into clusters, the number of clusters utilised, and the method by
which our sector/market indexes are constructed all impact the ability
of the model to capture dependence. Whereas \cite{Heinen2008} tested
their CAVA model using ten almost equally-sized industry sectors,
and \cite{Brechmann2013a} tested their RVMS model using five country-based
groupings of varying size, we have tested the CDCV model on a data
set from which a variable or fixed number of clusters may be formed
in each given time step. We have demonstrated that smaller cluster
sizes are preferable for capturing the majority of the bivariate rank
correlation, that these small clusters tend to introduce negative
dependence into the model and that this may be mitigated in practice
by the addition of a small perturbation to the index construction
process. While we have only performed a cursory analysis of which
clustering rules provide the best model performance, this approach
opens up a number areas for further analysis. Critically, we have
also shown that the performance of such models is not constant through
time. While this may seem an obvious conclusion it is an aspect of
these models that has not been fully addressed in the literature to
date.\\
\\
The applications of the CDCV model are significantly more diverse
than those of the market-sector models that it extends, due primarily
to its abstraction of the modelling framework from the underlying
data. While the CDCV model utilises the same hierarchical construction
used by \cite{Heinen2008,Brechmann2013a} and again recently used
by the Bi-factor copula model of \cite{Joe2014}, its primary contribution
is the inclusion of a clustering mechanism to render it applicable
to all data sets. This is a logical next step to the combination of
market-sector and factor-copula models, and continues the theme of
abstracting high-dimensional vine copula modelling frameworks from
data, as also pursued by \cite{year,Brechmann2014233,Krupskii:2013:FCM:2501262.2501499}.
This abstraction makes the CDCV model applicable to the analysis of
any set of non-independent variables, although we would expect it
to be most appropriate for data sets that are likely to exhibit clustering
in a number of dimensions, such as global stock portfolios. In such
a financial context, copulas are already used for a variety of purposes;
for example to model the dependence between stocks within a basket
or to model and simulate expected returns on portfolios of assets.
Computationally feasible vine copula models can in turn provide demonstrable
improvements over market standard approaches such as multivariate
copulas or simple covariance matrices. A purpose for which full vine
copula models have already been demonstrated to present such an improvement
is the optimisation of high dimensional stock portfolios. For example,
\cite{Low2013} showed that a CVaR-optimised portfolio with selection
based on a Clayton C-Vine copula model outperforms an equivalent multivariate
Clayton copula model for portfolios or 10 or more assets. While fully
implemented vine copulas were addressed by the authors, a natural
extension of our research would be to assess whether extensions of
traditional vine copula models (such as the CDCV model and the models
of \cite{year,Brechmann2014233,Joe2014}) provide sufficient accuracy
to be practically implemented for such portfolio optimisation.

%\bibliographystyle{plainnat}
%\addcontentsline{toc}{section}{\refname}\bibliography{My_Collection_THESIS_paper}
\bibliographystyle{plain}
\bibliography{Bibliography}

\appendix

%\section{Appendix}
\label{sec:Appendix}

\section{Definitions and Algorithms}

\subsection{Clustering Rules -- Distance Metrics\label{sub:Clustering-Rules--}}

\textcolor{black}{A selection of distance metrics commonly found in
the literature. Note that $\bar{x}_{i,j}$ is the mean $x_{i,j}$
value, A is the set of concordant pairs, B the set of discordant pairs,
and $y_{i,j}$ the set of ranked variables derived from the raw $x_{i,j}$
values.}

\begin{table}[H]
\begin{centering}
\textcolor{black}{\tiny{}}%
\begin{tabular*}{35pc}{@{\extracolsep{\fill}}llclllllll}
\hline 
 & \textcolor{black}{\scriptsize{Distance Metric ($D$)}} &  & \textcolor{black}{\scriptsize{$=d\left(x_{i},\, x_{j}\right)$, for
a pair of vectors $x_{i}$, $x_{j}$ each with $T$ time steps}} &  &  &  &  &  &  \\
\hline 
\hline 
 &  &  &  &  &  &  &  &  &  \\
 & \textcolor{black}{\scriptsize{Euclidean}} &  & \textcolor{black}{\scriptsize{$=\sqrt{\sum_{t=1}^{T}\left(\left(x_{i}\right)_{t}-\left(x_{j}\right)_{t}\right)^{2}}$}} &  &  &  &  &  &  \\
 &  &  &  &  &  &  &  &  &  \\
 & \textcolor{black}{\scriptsize{Manhattan}} &  & \textcolor{black}{\scriptsize{$=\sum_{t=1}^{T}\mid\left(x_{i}\right)_{t}-\left(x_{j}\right)_{t}\mid$}} &  &  &  &  &  &  \\
 &  &  &  &  &  &  &  &  &  \\
 & \textcolor{black}{\scriptsize{Pearson's-based}} &  & \textcolor{black}{\scriptsize{$=\sqrt{2 \cdot \left(1-\frac{\sum_{t=1}^{T}\left(\left(x_{i}\right)_{t}-\bar{x}_{i}\right)\left(\left(x_{j}\right)_{t}-\bar{x}_{j}\right)}{\sqrt{\sum_{t=1}^{T}\left(\left(x_{i}\right)_{t}-\bar{x}_{i}\right)^{2}\sum_{t=1}^{T}\left(\left(x_{j}\right)_{t}-\bar{x}_{j}\right)^{2}}}\right)}$}} &  &  &  &  &  &  \\
 &  &  &  &  &  &  &  &  &  \\
 & \textcolor{black}{\scriptsize{Kendalls Tau-based}} &  & \textcolor{black}{\scriptsize{$=\sqrt{2 \cdot \left(1-\frac{\sum_{t=1}^{T}1_{A}\left(x_{i},x_{j}\right)_{t}-\sum_{t=1}^{T}1_{B}\left(x_{i},x_{j}\right)_{t}}{\frac{1}{2}n(n-1)}\right)}$}} &  &  &  &  &  &  \\
 &  &  &  &  &  &  &  &  &  \\
 & \textcolor{black}{\scriptsize{Spearmans Rho-based}} &  & \textcolor{black}{\scriptsize{$=\sqrt{2 \cdot \left(1-\frac{\sum_{t=1}^{T}\left(\left(y_{i}\right)_{t}-\bar{y}_{i}\right)\left(\left(y_{j}\right)_{t}-\bar{y}_{j}\right)}{\sqrt{\sum_{t=1}^{T}\left(\left(y_{i}\right)_{t}-\bar{y}_{i}\right)^{2}\sum_{t=1}^{T}\left(\left(y_{j}\right)_{t}-\bar{y}_{j}\right)^{2}}}\right)}$}} &  &  &  &  &  &  \\
 &  &  &  &  &  &  &  &  &  \\
\hline 
\end{tabular*}
\par\end{centering}{\tiny \par}

\bigskip{}
\medskip{}

\centering{}Table A1 : Distance metrics for agglomerative clustering
\end{table}

\subsection{Clustering Rules -- Linkage Criterion\label{sub:Clustering-Rules---1}}

\textcolor{black}{This table highlights the most common linkage criteria
used in the literature, in addition to an Adapted Single criterion
that we have introduced. Note that $X_{p},\, X_{q}$ may be either
singleton elements or in-progress clusters of size $>1$, where the
set of all pairs of non-singleton clusters is denoted by $\Omega$,
the maximum cluster size is set by parameter $a$, the maximum
number of clusters is set by parameter $b$, and the resultant
number of clusters that would exist following a given linkage is denoted
by $X^{n}$.}

\begin{table}[H]
\begin{centering}
\textcolor{black}{\tiny{}}%
\begin{tabular*}{35pc}{@{\extracolsep{\fill}}llcll}
\hline 
 & \textcolor{black}{\scriptsize{Linkage Criterion}} &  & \textcolor{black}{\scriptsize{$=P\left(X_{p},\, X_{q}\right)$, to
be linked, given elements $x_{i}$, $x_{j}$ in clusters $X_{p},X_{q}$}} &  \\
\hline 
\hline 
 &  &  &  &  \\
 & \textcolor{black}{\scriptsize{Single}} &  & \textcolor{black}{\scriptsize{$=\underset{X_{p},X_{q}}{\min}\left\{ \underset{i,j}{\min}\left\{ d\left(x_{i},\, x_{j}\right)\,:\, x_{i}\in X_{p},\, x_{j}\in X_{q}\right\} \right\} $}} &  \\
 &  &  &  &  \\
 & \textcolor{black}{\scriptsize{Complete}} &  & \textcolor{black}{\scriptsize{$=\underset{X_{p},X_{q}}{\min}\left\{ \underset{i,j}{\max}\left\{ d\left(x_{i},\, x_{j}\right)\,:\, x_{i}\in X_{p},\, x_{j}\in X_{q}\right\} \right\} $}} &  \\
 &  &  &  &  \\
 & \textcolor{black}{\scriptsize{Average}} &  & \textcolor{black}{\scriptsize{$=\underset{X_{p},X_{q}}{\min}\left\{ \frac{1}{\mid X_{p}\mid\mid X_{q}\mid}\underset{x_{i}\in X_{p}}{\sum}\underset{x_{j}\in X_{q}}{\sum}d\left(x_{i},\, x_{j}\right)\right\} $}} &  \\
 &  &  &  &  \\
 & \textcolor{black}{\scriptsize{Adapted Single}} &  & \textcolor{black}{\scriptsize{$=\underset{X_{p},X_{q}}{\min}\left\{ \underset{i,j}{\min}\left\{ d\left(x_{i},\, x_{j}\right)\,:\, x_{i}\in X_{p},\, x_{j}\in X_{q}\right\} :,\,\left[X_{p},\, X_{q}\right]\notin\Omega,\, X_{p}<a,\, X_{q}<a,\, X^{n}\leq b\right\} $}} &  \\
 &  &  &  &  \\
\hline 
\end{tabular*}
\par\end{centering}{\tiny \par}

\bigskip{}
\medskip{}

\centering{}Table A2 : Linkage criteria for agglomerative clustering
\end{table}

\subsection{Clustering Algorithm -- Adapted Single\label{sub:Clustering-Algorithm--}}

\textcolor{black}{This pseudo code is for a general agglomerative
clustering algorithm. Resulting set of clusters is denoted $\Omega$,
where R is a pre-defined stopping rule, $l\left(\right)$ is the linkage
criterion and $d\left(\right)$ the distance metric. $L_{x,y}$ is
the pair of (possibly derived) time series selected from clusters
x and y by the linkage criterion, and $D_{x,y}$ is the distance calculated
between $L_{x,y}$. Finally, $\Vert\Omega\Vert$ is the size of the
set of all clusters.}

\renewcommand\arraystretch{0.9}

\begin{table}[H]
\begin{centering}
\tt
\textcolor{black}{\tiny{}}%
\begin{tabular*}{35pc}{@{\extracolsep{\fill}}l}
\hline
\textcolor{black}{\footnotesize{Algorithm to Cluster Assets}}  \\
\hline 
\hline 
 \\
\textcolor{black}{\footnotesize{\hspace{0.5cm}Select n asset time series $a_{i};\, i=1,...,n$}} \\
\textcolor{black}{\footnotesize{\hspace{0.5cm}$m=n$}} \\
\textcolor{black}{\footnotesize{\hspace{0.5cm}Select m clusters $C_{k}=\left\{ a_{i};\, a_{i}\in C_{k}\right\} ;\, k=1,...,m\,;\, i=1,...,n$}} \\
\textcolor{black}{\footnotesize{\hspace{0.5cm}$q=0$}} \\
\textcolor{black}{\footnotesize{\hspace{0.5cm}Set $R=FALSE$}} \\
\textcolor{black}{\footnotesize{\hspace{0.5cm}for $z\leftarrow 1,2,3,...$ }} \\
\textcolor{black}{\footnotesize{\hspace{0.5cm}\hspace{0.5cm}$\Omega=\left\{ C_{k};\, C_{k}\neq0\right\} $}} \\
\textcolor{black}{\footnotesize{\hspace{0.5cm}\hspace{0.5cm}Evaluate
stopping rule $R$}} \\
\textcolor{black}{\footnotesize{\hspace{0.5cm}\hspace{0.5cm}if $R=TRUE$
or $\Vert\Omega\Vert=1$ then}} \\
\textcolor{black}{\footnotesize{\hspace{0.5cm}\hspace{0.5cm}\hspace{0.5cm}Stop}} \\
\textcolor{black}{\footnotesize{\hspace{0.5cm}\hspace{0.5cm}else
if $R=FALSE$ then}} \\
\textcolor{black}{\footnotesize{\hspace{0.5cm}\hspace{0.5cm}\hspace{0.5cm}for
$x\leftarrow1,2,...,m+q$}} \\
\textcolor{black}{\footnotesize{\hspace{0.5cm}\hspace{0.5cm}\hspace{0.5cm}\hspace{0.5cm}for
$y\leftarrow1,2,...,m+q$}} \\
\textcolor{black}{\footnotesize{\hspace{0.5cm}\hspace{0.5cm}\hspace{0.5cm}\hspace{0.5cm}\hspace{0.5cm}if
$C_{x}=0$ or $C_{y}=0$ }} \\
\textcolor{black}{\footnotesize{\hspace{0.5cm}\hspace{0.5cm}\hspace{0.5cm}\hspace{0.5cm}\hspace{0.5cm}\hspace{0.5cm}$D_{x,y}=\infty^{+}$}} \\
\textcolor{black}{\footnotesize{\hspace{0.5cm}\hspace{0.5cm}\hspace{0.5cm}\hspace{0.5cm}\hspace{0.5cm}else}} \\
\textcolor{black}{\footnotesize{\hspace{0.5cm}\hspace{0.5cm}\hspace{0.5cm}\hspace{0.5cm}\hspace{0.5cm}\hspace{0.5cm}$L_{x,y}=l\left(C_{x},\, C_{y};\,\right)$ }} \\
\textcolor{black}{\footnotesize{\hspace{0.5cm}\hspace{0.5cm}\hspace{0.5cm}\hspace{0.5cm}\hspace{0.5cm}\hspace{0.5cm}$D_{x,y}=d\left(L_{x},\, L_{y}\right)$ }} \\
\textcolor{black}{\footnotesize{\hspace{0.5cm}\hspace{0.5cm}\hspace{0.5cm}\hspace{0.5cm}\hspace{0.5cm}end
if}} \\
\textcolor{black}{\footnotesize{\hspace{0.5cm}\hspace{0.5cm}\hspace{0.5cm}\hspace{0.5cm}end
for}} \\
\textcolor{black}{\footnotesize{\hspace{0.5cm}\hspace{0.5cm}\hspace{0.5cm}end
for}} \\
\textcolor{black}{\footnotesize{\hspace{0.5cm}\hspace{0.5cm}\hspace{0.5cm}$J_{x,y}=\left\{ \left[x,\, y\right];\, D_{x,y}=\min\left\{ D_{x,y}\right\} \right\} $}} \\
\textcolor{black}{\footnotesize{\hspace{0.5cm}\hspace{0.5cm}\hspace{0.5cm}$C_{m+1+q}=C_{J_{x}}+C_{J_{y}}$}} \\
\textcolor{black}{\footnotesize{\hspace{0.5cm}\hspace{0.5cm}\hspace{0.5cm}$q=q+1$}} \\
\textcolor{black}{\footnotesize{\hspace{0.5cm}\hspace{0.5cm}\hspace{0.5cm}$C_{J_{y}}=0$}} \\
\textcolor{black}{\footnotesize{\hspace{0.5cm}\hspace{0.5cm}\hspace{0.5cm}$C_{J_{x}}=0$}} \\
\textcolor{black}{\footnotesize{\hspace{0.5cm}\hspace{0.5cm}end
if}} \\
\textcolor{black}{\footnotesize{\hspace{0.5cm}end for}} \\
\textcolor{black}{\footnotesize{\hspace{0.5cm}$\Omega=\left\{ C_{k};\, C_{k}\neq0\right\} $}} \\
 \\
\hline 
\end{tabular*}
\par\end{centering}{\tiny \par}

\bigskip{}
\medskip{}

\centering{}Table A3 : Pseudo-code algorithm for agglomerative clustering
of assets
\end{table}

%\pagebreak{}

\subsection{Index Construction -- Example Rules\label{sub:Index-Construction--}}

\textcolor{black}{Index construction methods considered in Section
\ref{sub:Deriving-Hierarchical-Indexes} for a cluster of $n$ asset
timeseries of $T$ timesteps, where $m_{i}$ is the market capitalisation
of asset $i$ at a fixed point in time, $\tau_{i}^{t}$ is the sum
of Kendall's Tau values for all within-cluster bivariate pairs that
contain asset $i$, d is an arbitrarily defined parameter that increases
the severity of a given weighting, and $\sigma_{i}^{t}$ is the volatility
of asset $i$. We also define $X$ to be a matrix containing $n$
column vectors }$x_{1}^{t},...,x_{n}^{t}$ each of length equal to
the learn period used.

\begin{table}[H]
\begin{centering}
\textcolor{black}{\tiny{}}%
\begin{tabular*}{35pc}{@{\extracolsep{\fill}}llclllllll}
\hline 
 & \textcolor{black}{\scriptsize{Index}} &  & \textcolor{black}{\scriptsize{= $I\left(x_{1}^{t},...,x_{n}^{t}\right)$,
$\forall t\in\left(1,T\right)$}} &  &  &  &  &  &  \\
\hline 
\hline 
 &  &  &  &  &  &  &  &  &  \\
 & \textcolor{black}{\scriptsize{Simple Mean}} &  & \textcolor{black}{\scriptsize{$=\frac{\sum_{i=1}^{n}x_{i}^{t}}{n}$}} &  &  &  &  &  &  \\
 &  &  &  &  &  &  &  &  &  \\
 & \textcolor{black}{\scriptsize{``Market Capitalisation'' Weighted
Mean}} &  & \textcolor{black}{\scriptsize{$=\frac{\sum_{i=1}^{n}\left(m_{i}x_{i}^{t}\right)}{\sum_{i=1}^{n}m_{i}}$}} &  &  &  &  &  &  \\
 &  &  &  &  &  &  &  &  &  \\
 & \textcolor{black}{\scriptsize{``Sum of Kendall's Tau'' Weighted
Mean}} &  & \textcolor{black}{\scriptsize{$=\frac{\sum_{i=1}^{n}\left(\tau_{i}^{t}+d*\left(\tau_{i}^{t}-\min\left\{ \tau_{i}^{t}\right\} \right)x_{i}^{t}\right)}{\sum_{i=1}^{n}\left(\tau_{i}^{t}+d*\left(\tau_{i}^{t}-\min\left\{ \tau_{i}^{t}\right\} \right)\right)}$}} &  &  &  &  &  &  \\
 &  &  &  &  &  &  &  &  &  \\
 & \textcolor{black}{\scriptsize{``Volatility'' Weighted Mean}} &  & \textcolor{black}{\scriptsize{$=\frac{\sum_{i=1}^{n}\left(\sigma_{i}^{t}x_{i}^{t}\right)}{\sum_{i=1}^{n}\sigma_{i}^{t}}$}} &  &  &  &  &  &  \\
 &  &  &  &  &  &  &  &  &  \\
 & \textcolor{black}{\scriptsize{1st Principal Component}} &  & \textcolor{black}{\scriptsize{$=X.\left(\underset{\shortparallel w\shortparallel=1}{\arg\,\max}\left\{ \frac{w^{T}X^{T}Xw}{w^{T}w}\right\} \right)$}} &  &  &  &  &  &  \\
 &  &  &  &  &  &  &  &  &  \\
\hline 
\end{tabular*}
\par\end{centering}{\tiny \par}

\bigskip{}
\medskip{}

\centering{}Table A4 : Construction methods for sector and market
indexes
\end{table}

\subsection{C-Vine Model-fitting Algorithm\label{sub:C-Vine-Model-fitting-Algorithm}}

We provide below pseudo-code for a general C-Vine copula fitting algorithm
that utilises these $h$-functions, based on the algorithms provided
by \cite{Aas2006}. In this algorithm we obtain a vector $Q_{ik}\left(\Psi_{ik},\,\Theta_{ik},\, L_{ik}\right)$
of fitted bivariate copulas between the $n$ time series indexed via
$i$ and $k$, where each element $Q_{ik}$ consists of a selected
copula family $\Psi_{ik}$, and set of fitted parameters $\Theta_{ik}$
and a log-likelihood $L_{ik}$. These values are obtained by application
of the functions \emph{fitCopula( )} from the package \emph{\{copula\}}
and \emph{AIC( )} from the package \emph{\{stats\}}, which perform
the bivariate fitting process described in Section \ref{sec:The-StatVine-Model}
for time series already transformed via appropriate probability integral
transformations to the $U[0,1]$ space. \textcolor{black}{In this
model-fitting algorithm for a C-Vine copula we select between $I$
copula families by AIC during each iteration.}

\begin{table}[H]
\begin{centering}
\tt
\textcolor{black}{\tiny{}}%
\begin{tabular*}{35pc}{@{\extracolsep{\fill}}l}
\hline 
\textcolor{black}{\footnotesize{C-Vine Fitting}} \\
\hline 
\hline 
 \\
\textcolor{black}{\small{\hspace{0.5cm}Introduce $x_{i}$; $i=1,...,n$
time series vectors on $U[0,1]$ to be fitted}} \\
\textcolor{black}{\small{\hspace{0.5cm}for $i\leftarrow2,3,...,n$}} \\
\textcolor{black}{\small{\hspace{0.5cm}\hspace{0.5cm}for $k\leftarrow1,2,...,i\text{\textminus}1$}} \\
\textcolor{black}{\small{\hspace{0.5cm}\hspace{0.5cm}\hspace{0.5cm}for
$c\leftarrow1,2,...,I$}} \\
\textcolor{black}{\small{\hspace{0.5cm}\hspace{0.5cm}\hspace{0.5cm}\hspace{0.5cm}$q_{c_{ik}}\left(\psi_{c_{ik}},\,\theta_{c_{ik}},\, l_{c_{ik}}\right) = fitCopula\left(x_{i},\, x_{k};\,\psi_{c_{ik}}=c\right)$}} \\
\textcolor{black}{\small{\hspace{0.5cm}\hspace{0.5cm}\hspace{0.5cm}\hspace{0.5cm}$a_{c_{ik}} = AIC\left(\theta_{c_{ik}},\, l_{c_{ik}}\right)$}} \\
\textcolor{black}{\small{\hspace{0.5cm}\hspace{0.5cm}\hspace{0.5cm}end
for }} \\
\textcolor{black}{\small{\hspace{0.5cm}\hspace{0.5cm}\hspace{0.5cm}$C_{ik} = \left\{ c_{ik};\, AIC\left(\theta_{c_{ik}},\, l_{c_{ik}}\right)=\min\left[a_{c_{ik}}\right]\right\} $}} \\
\textcolor{black}{\small{\hspace{0.5cm}\hspace{0.5cm}\hspace{0.5cm}$Q_{ik}\left(\Psi_{ik},\,\Theta_{ik},\, L_{ik}\right) = q_{C_{ik}}$}} \\
\textcolor{black}{\small{\hspace{0.5cm}\hspace{0.5cm}\hspace{0.5cm}$x_{i} = h_{\Psi_{ik}}\left(x_{i},\, x_{k};\,\Theta_{ik}\right)$}} \\
\textcolor{black}{\small{\hspace{0.5cm}\hspace{0.5cm}end for }} \\
\textcolor{black}{\small{\hspace{0.5cm}end for }} \\
 \\
\hline 
\end{tabular*}
\par\end{centering}{\tiny \par}

\bigskip{}
\medskip{}

\centering{}Table A5 : Pseudo-code algorithm for fitting a full C-Vine
copula
\end{table}

\subsection{C-Vine Simulation Algorithm\label{sub:C-Vine-Simulation-Algorithm}}

To simulate from a C-Vine, we begin with a random sample $w_{i}$
of data for each of our asset return variables, and then iteratively
``un-condition'' the sample through each tree of the vine, applying
``inverse $h$-functions'' at each step as necessary to obtain the
value of the previous conditioning variable. This is essentially a
reverse version of our C-Vine fitting algorithm, and provides us with
a single sample from the vine copula, denoted by the vector $x$.
\textcolor{black}{The following simulation algorithm for a C-Vine
copula is per \cite{Aas2006}. As earlier, we have that $v_{i-j}$
denotes all $v_{i}$ but excluding $v_{j}$.}

\begin{table}[H]
\begin{centering}
\tt
\textcolor{black}{\tiny{}}%
\begin{tabular*}{35pc}{@{\extracolsep{\fill}}l}
\hline 
\textcolor{black}{\footnotesize{C-Vine Simulation}} \\
\hline 
\hline 
 \\
\textcolor{black}{\small{\hspace{0.5cm}Sample $w_{i}$; $i=1;...n$
independent uniform on $[0,1]$}} \\
\textcolor{black}{\small{\hspace{0.5cm}$x_{1}=v_{1,1}=w_{1}$}} \\
\textcolor{black}{\small{\hspace{0.5cm}for $i\leftarrow2,3,...,n$}} \\
\textcolor{black}{\small{\hspace{0.5cm}\hspace{0.5cm}$v_{i,1}=w_{i}$}} \\
\textcolor{black}{\small{\hspace{0.5cm}\hspace{0.5cm}for $k\leftarrow i\lyxmathsym{\textminus}1,i\lyxmathsym{\textminus}2,...,1$}} \\
\textcolor{black}{\small{\hspace{0.5cm}\hspace{0.5cm}\hspace{0.5cm}$v_{i,1}=h^{\lyxmathsym{\textminus}1}(v_{i,1},v_{k,k},\Theta_{k,i\lyxmathsym{\textminus}k})$}} \\
\textcolor{black}{\small{\hspace{0.5cm}\hspace{0.5cm}end for }} \\
\textcolor{black}{\small{\hspace{0.5cm}\hspace{0.5cm}$x_{i}=v_{i,1}$}} \\
\textcolor{black}{\small{\hspace{0.5cm}\hspace{0.5cm}if $i==n$
then }} \\
\textcolor{black}{\small{\hspace{0.5cm}\hspace{0.5cm}\hspace{0.5cm}Stop}} \\
\textcolor{black}{\small{\hspace{0.5cm}\hspace{0.5cm}end if}} \\
\textcolor{black}{\small{\hspace{0.5cm}\hspace{0.5cm}for $j\leftarrow1,2,...,i\lyxmathsym{\textminus}1$}} \\
\textcolor{black}{\small{\hspace{0.5cm}\hspace{0.5cm}\hspace{0.5cm}$v_{i,j+1}=h(v_{i,j},v_{j,j},\Theta_{j,i\lyxmathsym{\textminus}j})$ }} \\
\textcolor{black}{\small{\hspace{0.5cm}\hspace{0.5cm}end for }} \\
\textcolor{black}{\small{\hspace{0.5cm}end for}} \\
 \\
\hline 
\end{tabular*}
\par\end{centering}{\tiny \par}

\bigskip{}
\medskip{}

\centering{}Table A6 : Pseudo-code algorithm for simulating from a
full C-Vine copula
\end{table}

%\pagebreak{}

\subsection{CDCV Model-fitting Algorithm\label{sub:StatVine-Model-fitting-Algorithm}}

We provide an algorithm for fitting the CDCV Model in Table A7, where
$w_{E}$ is the sample market index, $w_{Ce}$ is the index for the
$e^{th}$ cluster, and $w_{z}^{Ce}$ is the $z^{th}$ sample asset
for the $e^{th}$ cluster. We denote the copula family fitted as $\psi$,
the fitted copula parameters by $\theta$, the log likelihood as $l$
and the corresponding AIC statistic $a$. The vector of selected copula
families is denoted $C$ and stored in $Q$ with parameters and log
likelihoods. In this notation there are $E$ clusters and $Z^{e}$
assets within each cluster. We choose from $I$ bivariate copula families
in each bivariate fitting and from $J$ multivariate copula families
in the joint-simplification process.

\begin{table}[H]
\begin{centering}
%\centering{}\textcolor{black}{\tiny{}}%
\tt
\textcolor{black}{\tiny{}}%
\begin{tabular*}{35pc}{@{\extracolsep{\fill}}ll}
\hline 
\textcolor{black}{\footnotesize{CDCV Fitting}} &  \\
\hline 
\hline 
 &  \\
\textcolor{black}{\small{\hspace{0.5cm}Introduce $w_{M}$ time series
vector on $U[0,1]$ for the market index}} &  \\
\textcolor{black}{\small{\hspace{0.5cm}Introduce $w_{Ce}$; $e=1,...,E$
time series vectors on $U[0,1]$ for the derived cluster indexes}} &  \\
\textcolor{black}{\small{\hspace{0.5cm}Introduce $w_{z}^{Ce}$; $z=1,...,Z^{e},\, e=1,...,E$
time series vectors on $U[0,1]$ for the assets}} &  \\
\textcolor{black}{\small{\hspace{0.5cm}Define $\psi_{c_{\left(\cdot,\cdot\right)}}$
as the fitted bivariate copula family for the copula $c$}} &  \\
\textcolor{black}{\small{\hspace{0.5cm}Define $\theta_{c_{\left(\cdot,\cdot\right)}}$
as the fitted bivariate copula parameter(s) for the copula $c$}} &  \\
\textcolor{black}{\small{\hspace{0.5cm}Define $l_{c_{\left(\cdot,\cdot\right)}}$
as the fitted bivariate copula log likelihood for the copula $c$}} &  \\
\textcolor{black}{\small{\hspace{0.5cm}Define $q_{c_{\left(\cdot,\cdot\right)}}$
as a vector containing fitted copula families, parameters and log
likelihoods}} &  \\
\textcolor{black}{\small{\hspace{0.5cm}Define $a_{c_{\left(\cdot,\cdot\right)}}$
as a vector containing AIC values}} &  \\
\textcolor{black}{\small{\hspace{0.5cm}Define simplified notation
for sector to market pairs: $\Lambda=\left(w_{Ce},w_{M}\right)$}} &  \\
\textcolor{black}{\small{\hspace{0.5cm}Define simplified notation
for asset to market pairs: $\Pi=\left(w_{z}^{Ce},w_{M}\right)$}} &  \\
\textcolor{black}{\small{\hspace{0.5cm}Define simplified notation
for asset to sector pairs: $\Omega=\left(w_{z}^{Ce},w_{Ce}\right)$}} &  \\
\textcolor{black}{\small{\hspace{0.5cm}Define simplified notation
for the set of all assets: $\diamondsuit=\left(w_{1}^{C_{1}},...,w_{Z^{1}}^{C_{1}},......,w_{1}^{C_{E}},...,w_{Z^{E}}^{C_{E}}\right)$}} &  \\
 &  \\
\textcolor{black}{\small{\hspace{0.5cm}\#\# Loop Through Clusters
\#\#}} &  \\
\textcolor{black}{\small{\hspace{0.5cm}for $e\leftarrow1,2,...,E$}} &  \\
\textcolor{black}{\small{\hspace{0.5cm}\hspace{0.5cm}\#\# Fit Cluster
Index to Market Index Copula \#\#}} &  \\
\textcolor{black}{\small{\hspace{0.5cm}\hspace{0.5cm}for $c\leftarrow1,2,...,I$}} &  \\
\textcolor{black}{\small{\hspace{0.5cm}\hspace{0.5cm}\hspace{0.5cm}$q_{c_{\Lambda}}\left(\psi_{c_{\Lambda}},\,\theta_{c_{\Lambda}},\, l_{c_{\Lambda}}\right) = fitCopula\left(w_{Ce},\, w_{M};\,\psi_{c_{\Lambda}}=c\right)$}} &  \\
\textcolor{black}{\small{\hspace{0.5cm}\hspace{0.5cm}\hspace{0.5cm}$a_{c_{\Lambda}} = AIC\left(\theta_{c_{\Lambda}},\, l_{c_{\Lambda}}\right)$}} &  \\
\textcolor{black}{\small{\hspace{0.5cm}\hspace{0.5cm}end for}} &  \\
\textcolor{black}{\small{\hspace{0.5cm}\hspace{0.5cm}$C_{\Lambda} = \left\{ c_{\Lambda};\, AIC\left(\theta_{c_{\Lambda}},\, l_{c_{\Lambda}}\right)=\min\left[a_{c_{\Lambda}}\right]\right\} $}} &  \\
\textcolor{black}{\small{\hspace{0.5cm}\hspace{0.5cm}$Q_{\Lambda}\left(\Psi_{\Lambda},\,\Theta_{\Lambda},\, L_{\Lambda}\right) = q_{C_{\Lambda}}$}} &  \\
\textcolor{black}{\small{\hspace{0.5cm}\hspace{0.5cm}$w_{Ce} = h_{\Psi_{\Lambda}}\left(w_{Ce},\, w_{M};\,\Theta_{\Lambda}\right)$}} &  \\
\textcolor{black}{\small{\hspace{0.5cm}\hspace{0.5cm}\#\# Loop Through
Assets \#\#}} &  \\
\textcolor{black}{\small{\hspace{0.5cm}\hspace{0.5cm}for $z\leftarrow1,2,...,Z^{e}$}} &  \\
\textcolor{black}{\small{\hspace{0.5cm}\hspace{0.5cm}\hspace{0.5cm}\#\#
Fit Asset to Market Index Copula \#\#}} &  \\
\textcolor{black}{\small{\hspace{0.5cm}\hspace{0.5cm}\hspace{0.5cm}for
$c\leftarrow1,2,...,I$}} &  \\
\textcolor{black}{\small{\hspace{0.5cm}\hspace{0.5cm}\hspace{0.5cm}\hspace{0.5cm}$q_{c_{\Pi}}\left(\psi_{c_{\Pi}},\,\theta_{c_{\Pi}},\, l_{c_{\Pi}}\right) = fitCopula\left(w_{z}^{Ce},\, w_{M};\,\psi_{c_{\Pi}}=c\right)$}} &  \\
\textcolor{black}{\small{\hspace{0.5cm}\hspace{0.5cm}\hspace{0.5cm}\hspace{0.5cm}$a_{c_{\Pi}} = AIC\left(\theta_{c_{\Pi}},\, l_{c_{\Pi}}\right)$}} &  \\
\textcolor{black}{\small{\hspace{0.5cm}\hspace{0.5cm}\hspace{0.5cm}end
for}} &  \\
\textcolor{black}{\small{\hspace{0.5cm}\hspace{0.5cm}\hspace{0.5cm}$C_{\Pi} = \left\{ c_{\Pi};\, AIC\left(\theta_{c_{\Pi}},\, l_{c_{\Pi}}\right)=\min\left[a_{c_{\Pi}}\right]\right\} $}} &  \\
\textcolor{black}{\small{\hspace{0.5cm}\hspace{0.5cm}\hspace{0.5cm}$Q_{\Pi}\left(\Psi_{\Pi},\,\Theta_{\Pi},\, L_{\Pi}\right) = q_{C_{\Pi}}$}} &  \\
\textcolor{black}{\small{\hspace{0.5cm}\hspace{0.5cm}\hspace{0.5cm}$w_{z}^{Ce} = h_{\Psi_{\Pi}}\left(w_{z}^{Ce},\, w_{M};\,\Theta_{\Pi}\right)$}} &  \\
\textcolor{black}{\small{\hspace{0.5cm}\hspace{0.5cm}\hspace{0.5cm}\#\#
Fit Asset to Cluster Index Copula \#\#}} &  \\
\textcolor{black}{\small{\hspace{0.5cm}\hspace{0.5cm}\hspace{0.5cm}for
$c\leftarrow1,2,...,I$}} &  \\
\textcolor{black}{\small{\hspace{0.5cm}\hspace{0.5cm}\hspace{0.5cm}\hspace{0.5cm}$q_{c_{\Omega}}\left(\psi_{c_{\Omega}},\,\theta_{c_{\Omega}},\, l_{c_{\Omega}}\right) = fitCopula\left(w_{z}^{Ce},\, w_{Ce};\,\psi_{c_{\Omega}}=c\right)$}} &  \\
\textcolor{black}{\small{\hspace{0.5cm}\hspace{0.5cm}\hspace{0.5cm}\hspace{0.5cm}$a_{c_{\Omega}} = AIC\left(\theta_{c_{\Omega}},\, l_{c_{\Omega}}\right)$}} &  \\
% &  \\
%\rm{\textit{Continued overleaf...}} &  \\
%\hline 
% &  \\
%\end{tabular*}
%\end{table}
%
%
%\begin{table}[H]
%\begin{centering}
%\hphantom{}
%\par\end{centering}
%
%\begin{centering}
%\tt
%\textcolor{black}{\tiny{}}%
%\begin{tabular*}{35pc}{@{\extracolsep{\fill}}ll}
%\hline 
%\rm{\textit{...from previous page.}} &  \\
% &  \\
\textcolor{black}{\small{\hspace{0.5cm}\hspace{0.5cm}\hspace{0.5cm}end
for}} &  \\
\textcolor{black}{\small{\hspace{0.5cm}\hspace{0.5cm}\hspace{0.5cm}$C_{\Omega} = \left\{ c_{\Omega};\, AIC\left(\theta_{c_{\Omega}},\, l_{c_{\Omega}}\right)=\min\left[a_{c_{\Omega}}\right]\right\} $}} &  \\
\textcolor{black}{\small{\hspace{0.5cm}\hspace{0.5cm}\hspace{0.5cm}$Q_{\Omega}\left(\Psi_{\Omega},\,\Theta_{\Omega},\, L_{\Omega}\right) = q_{C_{\Omega}}$}} &  \\
\textcolor{black}{\small{\hspace{0.5cm}\hspace{0.5cm}\hspace{0.5cm}$w_{z}^{Ce} = h_{\Psi_{\Omega}}\left(w_{z}^{Ce},\, w_{Ce};\,\Theta_{\Omega}\right)$}} &  \\
\textcolor{black}{\small{\hspace{0.5cm}\hspace{0.5cm}end for }} &  \\
\textcolor{black}{\small{\hspace{0.5cm}end for }} &  \\
\textcolor{black}{\small{\hspace{0.5cm}\#\# Fit Multivariate Copula
\#\#}} &  \\
\textcolor{black}{\small{\hspace{0.5cm}for $c\leftarrow1,2,...,J$}} &  \\
\textcolor{black}{\small{\hspace{0.5cm}\hspace{0.5cm}$q_{c_{\diamondsuit}}\left(\psi_{c_{\diamondsuit}},\,\theta_{c_{\diamondsuit}},\, l_{c_{\diamondsuit}}\right) = fitCopula\left(w_{1}^{C_{1}},...,w_{Z^{1}}^{C_{1}},......,w_{1}^{C_{E}},...,w_{Z^{E}}^{C_{E}};\,\psi_{c_{\diamondsuit}}=c\right)$}} &  \\
\textcolor{black}{\small{\hspace{0.5cm}\hspace{0.5cm}$a_{c_{\diamondsuit}} = AIC\left(\theta_{c_{\diamondsuit}},\, l_{c_{\diamondsuit}}\right)$}} &  \\
\textcolor{black}{\small{\hspace{0.5cm}end for}} &  \\
\textcolor{black}{\small{\hspace{0.5cm}$C_{\diamondsuit} = \left\{ c_{\diamondsuit};\, AIC\left(\theta_{c_{\diamondsuit}},\, l_{c_{\diamondsuit}}\right)=\min\left[a_{c_{\diamondsuit}}\right]\right\} $}} &  \\
\textcolor{black}{\small{\hspace{0.5cm}$Q_{\diamondsuit}\left(\Psi_{\diamondsuit},\,\Theta_{\diamondsuit},\, L_{\diamondsuit}\right) = q_{C_{\diamondsuit}}$}} &  \\
 &  \\
\hline 
\end{tabular*}
\par\end{centering}{\tiny \par}

\bigskip{}
\medskip{}

\centering{}Table A7 : Pseudo-code algorithm for fitting a CDCV copula
model
\end{table}

\subsection{CDCV Simulation Algorithm\label{sub:StatVine-Simulation-Algorithm}}

For the simulation algorithm we start by simulating the asset return
random variables from a multivariate copula rather than from a standard
uniform distribution. \textcolor{black}{In this simulation algorithm
for a CDCV copula model, $h_{\Psi_{\Omega}}^{-1}\left(\cdot\right)$
utilises the $Q_{\Omega}$ conditional fitting results and thus we
need not include an h-function. This is in line with the approach
of \cite{Heinen2008} but differs from the C-Vine algorithm in Table
A6 which assumes that families and parameters were obtained by fitting
unconditional copulas.}

\begin{table}[H]
\begin{centering}
\tt
\textcolor{black}{\tiny{}}%
\begin{tabular*}{35pc}{@{\extracolsep{\fill}}ll}
\hline 
\textcolor{black}{\footnotesize{CDCV Simulation}} &  \\
\hline 
\hline 
 &  \\
\textcolor{black}{\small{Load multivariate fitting output $Q_{\diamondsuit}\left(\Psi_{\diamondsuit},\,\Theta_{\diamondsuit},\, L_{\diamondsuit}\right)$}} &  \\
\textcolor{black}{\small{Load bivariate fitting output $Q_{i}\left(\Psi_{i},\,\Theta_{i},\, L_{i}\right)$
for $i=\left\{ \Lambda,\Pi,\Omega\right\} $}} &  \\
\textcolor{black}{\small{Sample $w_{i}$; $i=1;...n$ from multivariate
copula family $\Psi_{\diamondsuit}$ with parameters $\Theta_{\diamondsuit}$}} &  \\
 &  \\
\textcolor{black}{\small{\#\# Loop Through Clusters \#\#}} &  \\
\textcolor{black}{\small{for $e\leftarrow1,2,...,E$}} &  \\
\textcolor{black}{\small{\hspace{0.5cm}\#\# Loop Through Assets \#\#}} &  \\
\textcolor{black}{\small{\hspace{0.5cm}for $z\leftarrow1,2,...,Z^{e}$}} &  \\
\textcolor{black}{\small{\hspace{0.5cm}\hspace{0.5cm}\#\# ``Un-condition''
on the Cluster Index \#\#}} &  \\
\textcolor{black}{\small{\hspace{0.5cm}\hspace{0.5cm}$w_{z}^{Ce} = h_{\Psi_{\Omega}}^{-1}\left(w_{z}^{Ce},\, w_{Ce};\,\Theta_{\Omega}\right)$}} &  \\
\textcolor{black}{\small{\hspace{0.5cm}\hspace{0.5cm}\#\# ``Un-condition''
on the Market Index \#\#}} &  \\
\textcolor{black}{\small{\hspace{0.5cm}\hspace{0.5cm}$w_{z}^{Ce} = h_{\Psi_{\Pi}}^{-1}\left(w_{z}^{Ce},\, w_{Ce};\,\Theta_{\Pi}\right)$}} &  \\
\textcolor{black}{\small{\hspace{0.5cm}end for}} &  \\
\textcolor{black}{\small{end for}} &  \\
 &  \\
\hline 
\end{tabular*}
\par\end{centering}{\tiny \par}

\bigskip{}
\medskip{}

\centering{}Table A8 : Pseudo-code algorithm for simulating from
a CDCV copula model
\end{table}

\end{document}